\documentclass[floatfix,prx,twocolumn,letterpaper,lengthcheck,superscriptaddress,showpacs,amssymb,amsmath,amsfonts,aps,altaffilletter,nofootinbib,nopreprintnumbers,showpacs,longbibliography]{revtex4-1}

\usepackage{color}
\usepackage{hyperref}
\usepackage{amsmath}
\usepackage{multirow}
\usepackage{graphicx}
\usepackage{caption}
\usepackage{subcaption}
\usepackage{placeins}

\begin{document}

\title[]{The existence and stability of marginally trapped surfaces}

\author{Daniel Pook-Kolb} 
\affiliation{Max-Planck-Institut f\"ur Gravitationsphysik (Albert
  Einstein Institute), Callinstr. 38, 30167 Hannover, Germany}
\affiliation{Leibniz Universit\"at Hannover, 30167 Hannover, Germany}

\author{Ofek Birnholtz}
\affiliation{Center for Computational Relativity and Gravitation,
  Rochester Institute of Technology,
  170 Lomb Memorial Drive, Rochester, New York 14623, USA}

\author{Badri Krishnan} 
\affiliation{Max-Planck-Institut f\"ur Gravitationsphysik (Albert
  Einstein Institute), Callinstr. 38, 30167 Hannover, Germany}
\affiliation{Leibniz Universit\"at Hannover, 30167 Hannover, Germany}

\author{Erik Schnetter}
\affiliation{Perimeter Institute for Theoretical Physics, Waterloo, 
  ON N2L 2Y5, Canada}
\affiliation{Physics \& Astronomy Department, University of Waterloo,
  Waterloo, ON N2L 3G1, Canada}
\affiliation{Center for Computation \& Technology, Louisiana State
  University, Baton Rouge, LA 70803, USA}

\date{2018-11-23}

\begin{abstract}

  Marginally outer trapped surfaces (MOTSs, or marginal surfaces in
  short) are routinely used in numerical simulations of black hole
  spacetimes.  They are an invaluable tool for locating and
  characterizing black holes quasi-locally in real time while the
  simulation is ongoing.  It is often believed that a MOTS can behave
  unpredictably under time evolution; an existing MOTS can disappear,
  and a new one can appear without any apparent reason.  In this
  paper we show that in fact the behavior of a MOTS is perfectly
  predictable and its behavior is dictated by a single real parameter,
  the \emph{stability parameter}, which can be monitored during the
  course of a numerical simulation.  We demonstrate the utility of the
  stability parameter to fully understand the variety of marginal
  surfaces that can be present in binary black hole initial data.  We
  also develop a new horizon finder capable of locating very highly
  distorted marginal surfaces and we show that even in these cases,
  the stability parameter perfectly predicts the existence and
  stability of marginal surfaces.

\end{abstract}

\maketitle

\section{Introduction}
\label{sec:intro}

Numerical solutions of Einstein's equations (with and without matter)
are an important tool in gravitational wave astronomy (see
e.g. \cite{BaumgarteShapiroBook,ShibataBook,AlcubierreBook,Abbott:2016apu,Pretorius:2005gq,Campanelli:2005dd,
Baker:2005vv,Zlochower:2005bj,Healy:2014yta}).  In
situations without any symmetries or without any preferred background
solution, these numerical solutions are typically the most accurate
and are frequently used as benchmarks against which various
approximation methods can be tested.  In simulations of black hole
spacetimes, computing the gravitational wave signal is only one part
of the solution.  It is important to characterize the black holes as
well.  For example, in a binary black hole merger, one would like to
track the location and velocity of the black holes, and their physical
properties such as mass, angular momentum and higher multipole
moments.  Similarly, when the black holes merge and the final black
hole is formed, one would like to know when a common horizon forms,
and what the properties of the final black hole are.

In numerical simulations, one typically uses marginally trapped
surfaces to answer these questions.  Previous numerical methods for
locating marginal surfaces are described in
\cite{Cadez1974,Anninos:1994pa,Anninos:1996ez,Shoemaker:2000ye,Huq:2000qx,Thornburg:2003sf,Schnetter:2004mc}.
As we shall define in more detail later, these are closed two
dimensional surfaces with the topology of a sphere, in a timeslice.
They have the property that the outgoing light rays emanating from
them in the normal outward direction have vanishing expansion.  The
outermost such surface at a given time is called the apparent horizon.
Through the Penrose-Hawking singularity theorems, the presence of such
surfaces, together with various energy conditions, indicates the
presence of a singularity in the future
\cite{Penrose:1964wq,Hawking:1969sw}.  While not the main focus of
this paper, the notion of a MOTS also forms the starting point for the
study of quasi-local black hole horizons.  If a MOTS evolves smoothly
in time, we can consider the smooth 3-manifold $\mathcal{H}$ formed by
stacking up the MOTSs at different times.  Several important results
hold for this world tube $\mathcal{H}$ in different situations.  For
example, one can formulate the laws of black hole mechanics, define
multipole moments, obtain balance laws for the area and multipole
moments, and calculate black hole entropy in quantum gravity; see
e.g. \cite{Ashtekar:2004cn,Booth:2005qc,Gourgoulhon:2005ng,Hayward:2004fz,Jaramillo:2011zw}.
The applications of most relevance for us are in numerical relativity
where they are regularly used to assign mass, angular momentum and
higher multipole moments to black holes (see
e.g. \cite{Dreyer:2002mx,Schnetter:2006yt,Gupta:2018znn}).

The alternative to MOTS and quasi-local horizons are event horizons.
Many of the classic results of black hole physics, such as the area
increase law, black hole thermodynamics etc. were initially based on
event horizons (see
e.g. \cite{Bekenstein:1973ur,Bekenstein:1974ax,Bekenstein:1977mv,Bardeen:1973gs,Hawking:1971tu}).
However, the global and teleological nature of event horizons are well
known.  For both practical and theoretical reasons, event horizons are
not ideal for extracting the parameters of black holes.  First, their
teleological nature makes it impossible to locate them in real time in
a numerical simulation (or an experiment); we need to know the entire
spacetime.  Furthermore, it is primarily in perturbative situations
that event horizons are useful for defining and extracting parameters
like black hole mass, spin etc. In more general situations far from
stationarity, such as during a binary black hole merger, event
horizons are not suitable for this purpose.  For example, it is easy
to construct examples where event horizons form and grow in flat
Minkowski space where there should be no flux of gravitational
radiation.  These issues are discussed in more detail elsewhere (see
e.g. \cite{Ashtekar:2004cn,Booth:2005qc,Faraoni:2015pmn,Visser:2014zqa,Hayward:2000ca,Cabero:2017avf}).

Despite the utility of marginal surfaces, there still remain many
doubts about their behavior under time evolution.  It is observed that
apparent horizons can jump discontinuously.  This jump is now well
understood to arise from the outer-most condition for apparent
horizons, and the underlying marginal surfaces are observed to evolve
smoothly.  It is still not clear if this would also hold for highly
distorted marginal surfaces.  Do highly distorted marginal surfaces
still continue smoothly?  Is there a quantifiable way in which one can
say that a more distorted MOTS is more unstable?  There are cases when
a marginal surface can no longer be found.  Is this due to problems
with the horizon finders or is it really that the marginal surface has
ceased to exist?  Apart from their intrinsic interest, answers to such
questions are important to ensure the reliability of various physical
quantities that are routinely calculated in numerical simulations.

\begin{figure*}
  \begin{subfigure}[b]{0.48\textwidth}
    \centering    
    \includegraphics[width=\textwidth]{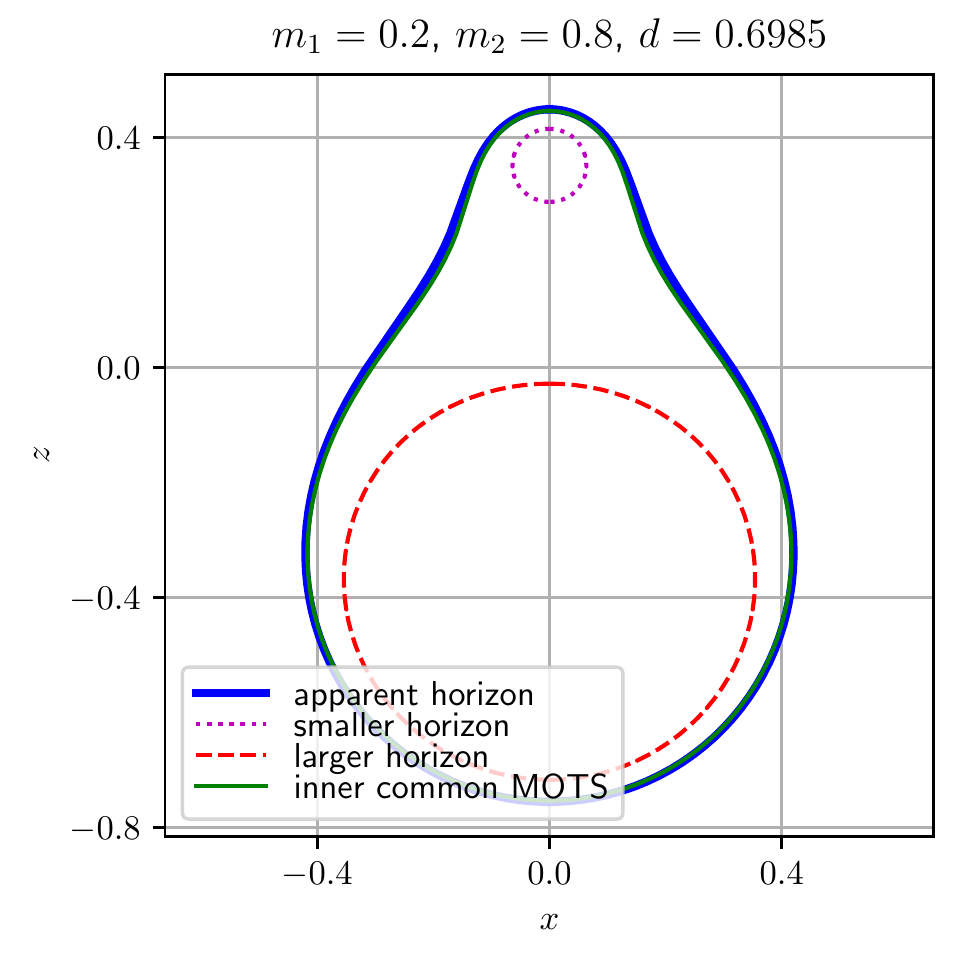}
    \caption{The various marginal surfaces shortly after the common MOTS is
      formed. For this separation, the inner common MOTS and the AH
      are very close to each other.}
    \label{fig:sub1}
    \centering
  \end{subfigure}
  \hfill
  \begin{subfigure}[b]{0.48\textwidth}
    \centering
    \includegraphics[width=\textwidth]{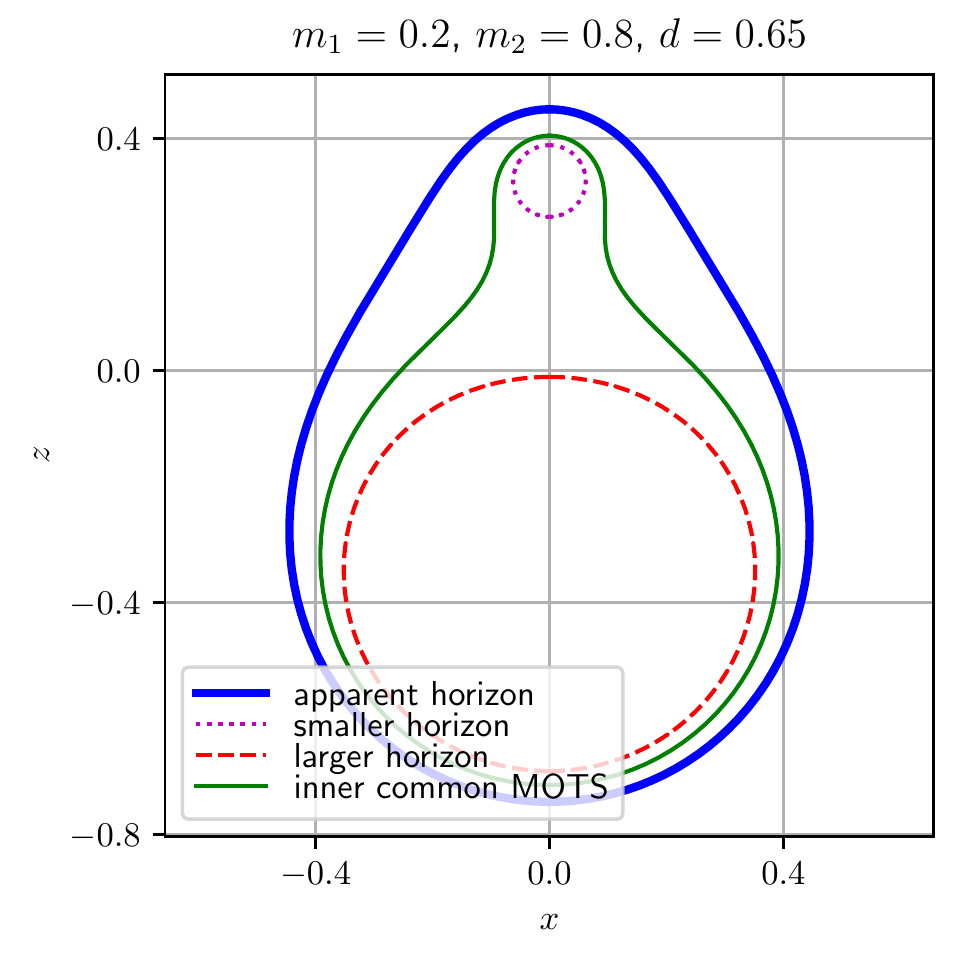}
    \caption{The inner common MOTS and the AH rapidly move away from
      each other as $d$ is decreased. The individual MOTSs are
      relatively undistorted at this stage. }
    \label{fig:sub2}
    \centering
  \end{subfigure}
  \hfill
  \begin{subfigure}[b]{0.48\textwidth}
    \centering
    \includegraphics[width=\textwidth]{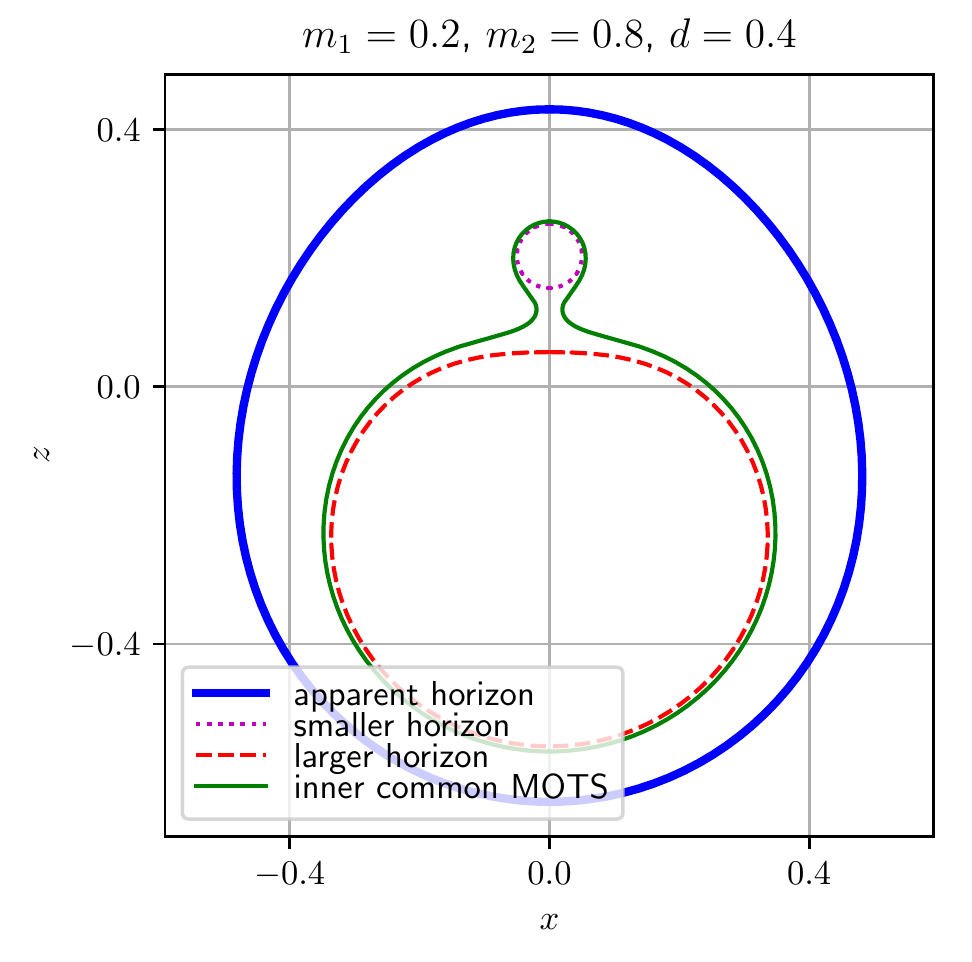}
    \caption{Here $d$ is further decreased and the inner common MOTS
      starts getting distorted while the AH becomes more uniform. The
      larger individual MOTS starts getting distorted as well. }
    \label{fig:sub3}
    \centering
  \end{subfigure}
  \hfill
  \begin{subfigure}[b]{.48\textwidth}
    \centering
    \includegraphics[width=\textwidth]{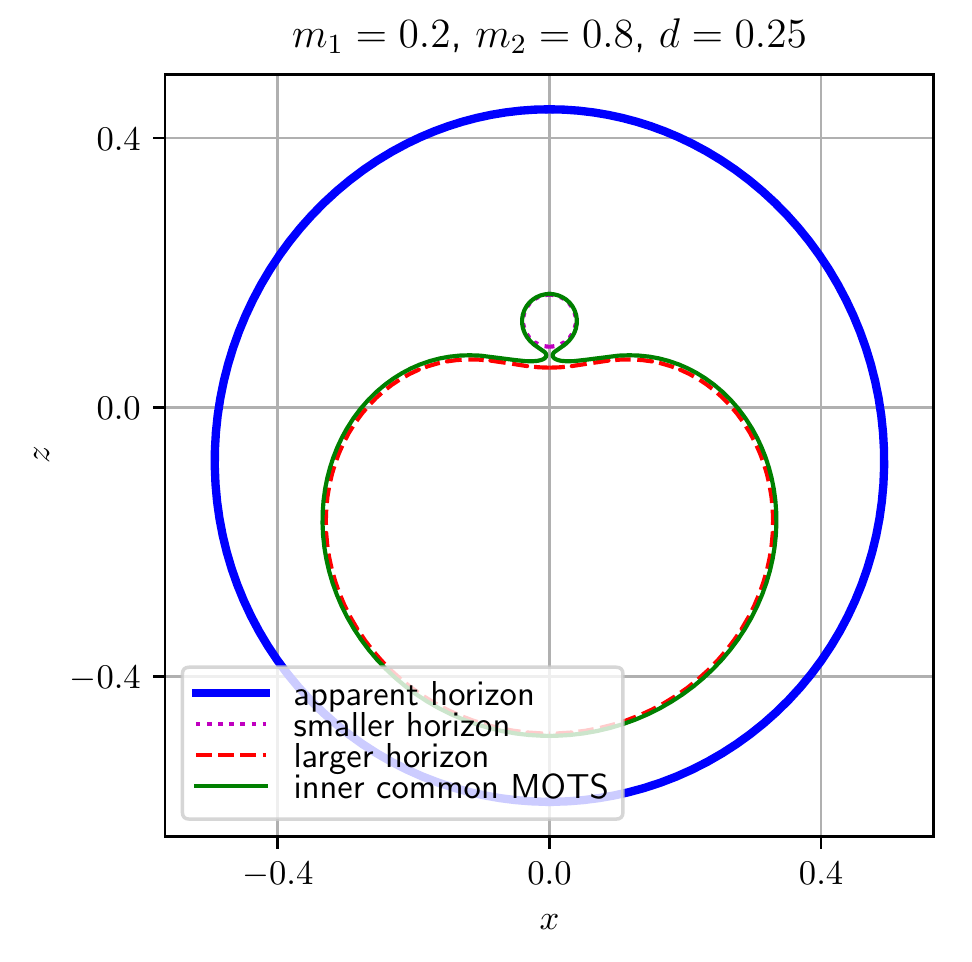}
    \caption{The inner common MOTS and the larger individual MOTS are
      now highly distorted; the inner common MOTS is clearly not star-shaped. The smaller individual MOTS acts as a
      barrier and prevents them from merging.}
    \label{fig:sub4}
    \centering
  \end{subfigure}
  
  \caption{Sequence for MOTS for BL data.  In each of the plots the
    masses are fixed at $m_1 = 0.2$, $m_2 = 0.8$, and
    the separation $d$ is successively smaller.  }
  \label{fig:brill_lindquist1}
\end{figure*}

At any given time, there can exist several marginally trapped surfaces
in a binary black hole spacetime.  This may include the two related
with the two individual black holes, and possibly two more, related to
the common final black hole (if it has formed).  The outermost of
these is called the apparent horizon but the other marginally trapped
surfaces are of interest as well (see e.g. Fig. \ref{fig:sub2} below
for an example.)  At late times, the apparent horizon will usually
approach the event horizon (see however
\cite{Williams:2007tp,Williams:2010pj}) but the fate of the other
marginal surfaces in the interior is not yet fully understood.  This
question is also of relevance for understanding the ``issue of the
final state''.  Mathematically this refers to the question of
non-linear stability of Kerr black holes.  Astrophysically,
restricting ourselves to binary systems, it refers to the fact that at
late times, irrespective of the initial configuration, the end state
is a Kerr black hole in equilibrium.  The exterior spacetime, and how
the gravitational wave signal shows this approach to equilibrium, even
how the horizon approaches equilibrium have been previously studied.
The structure of the interior is unresolved and the behavior of the
interior marginal surfaces is much less studied.

It might be argued that understanding the interior spacetime is of no
physical interest since this region is causally disconnected from the
external world where observations can be made - but this is
incorrect. Physical phenomena in the interior and in the exterior
regions are both a result of dynamics and non-linearities occurring
outside the event horizon.  Thus, we expect the two regions to be
correlated and in fact, such correlations have been shown to exist
\cite{Jaramillo:2011re,Jaramillo:2011rf,Jaramillo:2012rr,Rezzolla:2010df,Gupta:2018znn}.
The existence of these correlations leads to the interesting
possibility of inferring properties of the interior spacetime from
gravitational wave observations.  Thus, even from an astrophysical
perspective, it becomes important to understand the interior spacetime
in detail.

Mathematically, the conditions under which a marginal surface evolves
smoothly are known. Andersson et al.
\cite{Andersson:2005gq,Andersson:2007fh,Andersson:2008up,Booth:2006bn}
have shown that the key object for understanding these issues is the
\emph{stability operator}. This is a second order elliptic, possibly
non-self adjoint differential operator defined on every MOTS.  If its
smallest eigenvalue (which is guaranteed to be real) is positive, then
the MOTS evolves smoothly in time. Much less is known rigorously for
the cases when the principal eigenvalue is negative. In these cases,
the world tube swept out by the MOTSs is not spacelike; this is
studied in \cite{Bousso:2015qqa,Bousso:2015mqa}.  To our knowledge,
this stability operator has never been used in a numerical calculation
so far.  In this paper we demonstrate the utility of the stability
operator in numerical relativity, even for unstable and extremely
distorted marginal surfaces.

In order to really test the utility of the stability operator, we need
to calculate it for marginal surfaces which are extremely distorted.
The difficulty is that current horizon finders are generally not
capable of locating highly distorted surfaces. The most commonly used
numerical algorithms make certain assumptions on the marginal surfaces
they are trying to locate \cite{Thornburg:2006zb}.  For example, the
\texttt{AHFinderDirect} method available in the Einstein Toolkit
\cite{Loffler:2011ay,EinsteinToolkit:web,Thornburg:2003sf} assumes
that the surface can be represented by a single valued function
$h(\theta,\phi)$ of the usual angles $(\theta,\phi)$ in some spherical
coordinate system.  When this condition is satisfied, then the
marginal surface equation can be cast as an elliptic equation for
$h(\theta,\phi)$ which can be solved efficiently.  This condition
requires that any ray drawn from the origin of coordinates intersects
the surface exactly once, and a surface satisfying this condition is
said to be \emph{star-shaped}.  It is not expected that marginal
surfaces should always satisfy this condition, and we shall study
explicit counter-examples below.  The second goal of this paper is
thus the development and implementation of a new numerical method for
finding marginal surfaces which is computationally as fast as
\texttt{AHFinderDirect}, yet is capable of finding arbitrarily
distorted surfaces.

Armed with our new horizon finder, we investigate with high numerical
precision the various marginal surfaces which can exist in a simple
binary black hole initial data set, namely the Brill-Lindquist (BL)
data set representing non-spinning black holes at a moment of time
symmetry.  A similar project was first initiated in
\cite{Jaramillo:2009zz} applied to a different initial data
construction (Bowen-York), however only limited results, for the equal
mass case, were presented there.  The MOTS finder used in that work
was a pseudo-spectral axisymmetric code using bi-spherical coordinates
and therefore not easily generalizable.  See also
\cite{Hussain:2017ihw} for a study of distorted horizons in extreme
mass ratio systems.  Despite the simplicity of BL initial data, we
demonstrate the remarkably rich behavior of marginal surfaces, shown
here in such great detail for the first time.\footnote{Read the
  previous sentence in David Attenborough's voice.} The MOTSs in BL
data do have some special features which we shall mention later.
However, as far as the goals of our study are concerned, nothing is
lost by restricting ourselves to BL data. Just as for generic initial
data, the number of marginal surfaces is the same, and these surfaces
are not any less distorted.  Furthermore, since numerical accuracy is
of paramount importance here, it is more fruitful to initially focus
on studying sequences of initial data where additional numerical
errors due to time evolution can be ignored.  Once the basic link
between stability and existence has been established, as will be done
in this paper, we can proceed to apply these methods to other initial
data, and more importantly, to time evolutions in forthcoming work.

The plan for the rest of this paper is as follows.  The basic
definitions and mathematical properties of marginally trapped surfaces
are given in Sec.~\ref{subsec:definitions}, and
Sec.~\ref{subsec:numerics} summarizes existing methods for locating
marginal surfaces.  Sec.~\ref{sec:motsfinder} describes our new
method, including the parameterization in
Sec.~\ref{subsec:parameterization}, the algorithm in
Sec.~\ref{subsec:algorithm}, and its validation in
Sec.~\ref{subsec:convergence}.  Sec.~\ref{sec:brill} provides the
first set of results, by applying this method to a simple binary black
hole initial data, namely the BL data, representing the
head-on collision of two non-spinning black holes initially at rest,
considering various values of the mass-ratio and various separations.
Sec.~\ref{sec:mainresults} explains many of the results seen in
Sec.~\ref{sec:brill} in terms of stability.  It shows the crucial link
between the existence and stability of marginal surfaces.  Finally,
the universal behavior of the apparent horizon as $d\rightarrow 0$ is
shown in Sec.~\ref{sec:multipoles}, and final conclusions are given in
Sec.\ref{sec:conclusions}.

\section{Locating marginally trapped surfaces}
\label{sec:motsintro}

\subsection{Definitions and properties}
\label{subsec:definitions}

Let $\mathcal{S}$ be a closed spacelike 2-surface embedded in a
spacetime $\mathcal{M}$ with a Lorentzian metric $g_{ab}$ of signature
$({-}\,{+}\,{+}\,{+})$.  Let $q_{ab}$ be the Riemannian metric on $\mathcal{S}$
obtained by restricting $g_{ab}$ to vectors tangent to $\mathcal{S}$.
At any point $p \in \mathcal{S}$, we can perform an orthogonal
decomposition of the tangent space
$T_p\mathcal{M} = T_p\mathcal{S}\oplus T_p\mathcal{S}^\perp$.  The
orthogonal space $T_p\mathcal{S}^\perp$ has a Lorentzian metric, and
we can choose a basis of future directed null vectors $(\ell^a, n^a)$
in this vector space.  We assume also that it is possible to assign an
\emph{outward} direction on $\mathcal{S}$, and by convention $\ell^a$ is
taken to be outward pointing and $n^a$ inward pointing.  We can in
principle rescale $\ell^a$ and $n^a$ by positive definite functions,
however it is convenient to tie the scalings of $(\ell^a,n^a)$
together so that their inner product is preserved: $\ell\cdot n = -1$.
We are then left with possible rescalings (the boost transformations)
such that $\ell^a\rightarrow f\ell^a$, $n^a\rightarrow f^{-1}n^a$,
$f>0$.

Let $\nabla_a$ be the derivative operator on $\mathcal{M}$ compatible
with $g_{ab}$.  The expansions of $\ell^a$ and $n^a$ are respectively
\begin{equation}
  \Theta_{(\ell)} = q^{ab}\nabla_a\ell_b\,,\quad \Theta_{(n)} = q^{ab}\nabla_an_b\,.
\end{equation}
Under a boost, the expansions scale as
$\Theta_{(\ell)} \rightarrow f\Theta_{(\ell)} $, and
$\Theta_{(n)} \rightarrow f^{-1}\Theta_{(n)}$.

$\mathcal{S}$ is said to be a marginally outer trapped surface (MOTS,
or simply marginal surface) if $\Theta_{(\ell)} = 0$.
This definition is boost invariant (and invariant under all Lorentz
transformations which preserve the direction of $\ell^a$).  While we
will not generally require any condition on $\Theta_{(n)}$, it will be
negative in most physical situations, and it is important to check
that this is indeed the case.  Closed spacelike 2-surfaces with
$\Theta_{(\ell)} = 0$ and $\Theta_{(n)}<0$ are called marginally
future trapped surfaces.  Even in Schwarzschild spacetime, there are
examples of non-symmetric spatial slices which get arbitrarily close
to the future singularity, but do not in fact contain any marginally
trapped surfaces.  These slices would have marginal surfaces,
i.e. with $\Theta_{(\ell)}=0$, but they would not satisfy
$\Theta_{(n)}<0$ \cite{Wald:1991zz,Schnetter:2005ea}.

In the situation of interest for us, namely numerical simulations of
Einstein's equations as an initial value problem, the spacetime
$\mathcal{M}$ is foliated by spacelike Cauchy surfaces $\Sigma_t$
labeled by a real parameter $t$. Let $\mathcal{S}_t$ be a MOTS in
$\Sigma_t$.  Further, $t^a$ denotes the unit timelike normal to
$\Sigma_t$, $r^a$ the unit spacelike normal to $\mathcal{S}$ in
$\Sigma_t$, $h_{ab}$ the Riemannian metric on $\Sigma_t$ induced by
$g_{ab}$, and $D_a$ the derivative operator on $\Sigma_t$ compatible
with $h_{ab}$.  Denote the extrinsic curvature of $\Sigma_t$ by
$K_{ab}:=- h_{a}^ch_b^d\nabla_ct_d$ (the negative sign is conventional
in the numerical relativity literature).  A suitable choice of the
null normals is
\begin{equation}
  \ell^a = \frac{1}{\sqrt{2}} \left(t^a+ r^a\right)\,,\quad n^a = \frac{1}{\sqrt{2}}\left(t^a - r^a\right)\,.
\end{equation}
Note that for a given $\mathcal{S}_t$ choosing a different $\Sigma_t$
corresponds to some boost transformation. The MOTS condition is, as we
have seen, boost invariant. Thus, if two hypersurfaces $\Sigma_t$ and
$\Sigma_t^\prime$ contain $\mathcal{S}$, then $\mathcal{S}$ is a MOTS
in both hypersurfaces.

In terms of $t^a$ and $r^a$, $\Theta_{(\ell)}=0$ is equivalent to
\begin{equation}
\label{eq:motsequation}
  D_ar^a + K_{ab}r^ar^b - K = 0\,.
\end{equation}
This is the equation that must be solved on $\Sigma_t$, with given
$K_{ab}$ and $h_{ab}$, to locate the MOTS.  If we write the MOTS
$\mathcal{S}$ as the level-set of some function $F$, then
$r_a \propto D_aF$, and we get a second-order differential equation
for $F$.  We shall write this explicitly below, but for now we
conclude this section with a short summary of some properties of
MOTSs, and the relation between MOTSs and black hole horizons.

In the context of a time evolution, perhaps the first basic question
that arises is whether a given MOTS evolves smoothly.  It is \emph{a
  priori} possible that a MOTS is poorly behaved under time evolution
and may arbitrarily cease to exist, or a new MOTS may be formed
arbitrarily.  In numerical simulations, it is empirically observed
that a MOTS does behave smoothly under time evolution.  Mathematically
this has so far been proven for MOTSs with a physically well motivated
stability condition
\cite{Andersson:2005gq,Andersson:2007fh,Andersson:2008up,Booth:2006bn};
see also \cite{Booth:2017fob} for a study of variations of unstable
horizons.

Define first the \emph{variation} of $\Theta_{(\ell)}$ in the radial
direction $r^a$, $\delta_{fr}\Theta_{(\ell)}$ \cite{Newman1987}.  Let
$\mathcal{S}_\zeta$ be a smooth one-parameter family of closed
spacelike 2-surfaces on $\Sigma_t$ such that $\mathcal{S}_{\zeta=0}$
coincides with $\mathcal{S}$.  Each point $p$ on $\mathcal{S}$ traces
a smooth curve as $\zeta$ is varied. Let $k^a$ be the tangent vector
to this curve.  On each $\mathcal{S}_\zeta$ calculate the outgoing
null expansion $\Theta_{(\ell)}^\zeta$ and differentiate it with
respect to $\zeta$. This yields the variation of $\Theta_{(\ell)}$
along $k^a$:
\begin{equation}
  \delta_k\Theta_{(\ell)} := \left.\frac{d\Theta_{(\ell)}^\zeta}{d\zeta } \right|_{\zeta = 0}\,.
\end{equation}
It easy to see that the variation is linear for constants, but not for
a function: $\delta_{ck}\Theta_{(\ell)} = c\delta_{k}\Theta_{(\ell)}$
for a constant $c$, but
$\delta_{\psi k}\Theta_{(\ell)} \neq \psi\delta_{k}\Theta_{(\ell)}$
when $\psi$ is a non-constant function.

Choose $k^a = fr^a$, and define the operator $L$ as:
\begin{equation}
  \delta_{fr}\Theta_{(\ell)} =: Lf\,.
\end{equation}
It can be shown that $L$ is generally of the form
\begin{equation}
\label{eq:stability_operator}
  Lf = -\Delta f + \gamma^a \partial_a f + \beta f\,,
\end{equation}
where $\Delta$ is the Laplacian compatible with $q_{ab}$, $\gamma^a$
is some vector field on $\mathcal{S}$ (related to the angular
momentum) and $\beta$ is a scalar.  In the case of time symmetric data
(where the angular momentum and hence $\gamma^a$ vanishes), the
stability operator can be simplified to \cite{Andersson:2007fh}:
\begin{equation}
  \label{eq:stability_operator_bl}
  Lf = -\Delta f - (R_{ab}r^ar^b + \mathcal{K}_{ab}\mathcal{K}^{ab})f\,.
\end{equation}
Here $R_{ab}$ is the intrinsic Ricci tensor of $\Sigma_t$, and
$\mathcal{K}_{ab}$ is the second fundamental form of $\mathcal{S}$
embedded in $\Sigma_t$.  In this case $L$ is seen to be self adjoint.

In general however, from Eq.~(\ref{eq:stability_operator}), $L$ is a
second order elliptic operator with a discrete spectrum, which is
however not necessarily self-adjoint.  Its smallest eigenvalue, known
as the principal eigenvalue $\Lambda_0$, turns out to be real.  It was
shown that if the principal eigenvalue is positive, then the MOTS
evolves smoothly in time
\cite{Andersson:2005gq,Andersson:2007fh,Andersson:2008up}. This
stability condition is equivalent to saying that an outward
deformation of $\mathcal{S}$ makes it untrapped which is what we
expect to happen for the apparent horizon. We shall study the
stability operator and its first eigenvalues below in much greater
detail in specific examples.

In the introduction, we mentioned briefly the idea of stacking up
marginal surfaces at different times to construct a smooth 3-surface, a
quasi-local horizon.  The only notion from the formalism of
quasi-local horizons that we will use in this paper is that of mass
multipole moments.  Given a 2-sphere $\mathcal{S}$ with an axial
symmetry vector $\phi^a$ and an intrinsic Ricci scalar $\mathcal{R}$,
it turns out to be possible to construct a set of geometric multipole
moments which capture the intrinsic horizon geometry
\cite{Ashtekar:2004gp}.  Since we shall deal with time symmetric
situations, we do not need to consider the current multipole moments
and we restrict ourselves to the mass multipoles $I_n$.  These are
\begin{equation}
  \label{eq:massmultipoles}
  I_n = \sqrt{\frac{2n + 1}{4\pi}}\int_\mathcal{S} \mathcal{R} P_n(\zeta) d^2V 
\end{equation}
with the coordinate $\zeta \in [-1,1]$ (the analog of $\cos\theta$ in
standard spherical coordinates) defined as
\begin{equation}
  \partial_a\zeta = \frac{1}{R^2}\epsilon_{ba}\phi^b\,,\quad \int_{\mathcal{S}}\zeta d^2V = 0\,.
\end{equation}
Here $R$ is the radius of $\mathcal{S}$, $\epsilon$ the volume 2-form
on $\mathcal{S}$, and $P_n$ the $n^{\rm th}$ Legendre polynomial.

\subsection{Locating Star-shaped MOTSs numerically}
\label{subsec:numerics}

Most MOTS finders assume, first of all, that the 2-surface $\mathcal{S}$ is
topologically a sphere.  This is not a strong restriction as it can be
shown that generically the topology must be spherical, it is toroidal
only in very special situations, and higher genus surfaces are not
allowed \cite{Ashtekar:2003hk}.  The other common assumption is that
$\mathcal{S}$ demarcates a star-shaped region (Strahlk\"orper),
i.e. any ray drawn from the origin intersects $\mathcal{S}$ exactly
once. This is obviously a coordinate dependent definition since it
depends on the origin and on the set of coordinates in which straight
lines are drawn.  It excludes surfaces of the form shown in Fig. 1.1
of \cite{Thornburg:2006zb}, and we shall show explicit examples
of non-star shaped surfaces very much like it (e.g. Fig. \ref{fig:sub4}).  Given this assumption,
we can parameterize $\mathcal{S}$ as
\begin{equation}
  r = h(\theta,\phi) \,,
\end{equation}
where $r$ is the Euclidean distance from the origin of coordinates to
a point on $\mathcal{S}$, and $(\theta,\phi)$ are angular coordinates.
Here $(r,\theta,\phi)$ is the coordinate system used in the numerical
calculations, and we shall use mid-alphabet Latin indices $i,j,\ldots$ for tensors
in this coordinate system.  Thus, $h_{ij}$ is a Riemannian metric
(not to be confused with $h(\theta,\phi)$) and $K_{ij}$ the extrinsic
curvature.

We look for level sets of the function
\begin{equation}
  \label{eq:starshaped}
  F(r,\theta,\phi) = r - h(\theta,\phi)\,.
\end{equation}

The surfaces of constant $F$ define a sequence of surfaces and to
compute the normal, we start with the gradient of $F$:
\begin{equation}
  s_i = \partial_i F \,, \quad dF = dr - h_\theta d\theta - h_\phi d\phi\,.  
\end{equation}
The unit-normal $r_i$ is then
\begin{equation}
  r_i = \frac{s_i}{||s||}\,,\quad ||s||^2 = h^{ij}s_is_j\,.
\end{equation}
Since $s_i$ is directly related to the derivative of $h(\theta,\phi)$,
as opposed to $r_i$ which has a complicated non-linear dependence, it
is convenient to separate out the norm of $s_i$ in the expansion.
Looking at the three terms in Eq.~\eqref{eq:motsequation}, we see
immediately that $K_{ij}r^ir^j$ is proportional to $||s||^{-2}$.
Since $r^i$ is proportional to $||s||^{-1}$,
in the first term $D_ir^i$ it is clear that differentiation will lead to two
terms; one proportional to $||s||^{-3}$ and the other to
$||s||^{-1}$:
\begin{equation}
  D_i\left(\frac{s^i}{\sqrt{s\cdot s}} \right) = \frac{D_is^i}{(s\cdot s)^{1/2}} - \frac{s^is^jD_is_j}{(s\cdot s)^{3/2} }\,.
\end{equation}
Thus, the expansion of any level-set surface of $F$
is of the form
\begin{align}\label{eq:expansion_AB}
    \Theta = \frac{A}{||s||^{3}} + \frac{B}{||s||} + \frac{K_{ij}s^is^j}{||s||^2} - K,
\end{align}
where
\begin{subequations}\label{eq:AB_definition}
    \begin{align}
        \label{eq:AB_definition_A}
        A &:= - s^i s^j \partial_i s_j - \frac{1}{2} s^i (\partial_i h^{kl}) s_k s_l \,,\\
        \label{eq:AB_definition_B}
        B &:= (\partial_i h^{ij}) s_j + h^{ij} \partial_i s_j + (\partial_i \ln\sqrt{h}) s^i\,.
    \end{align}
\end{subequations}
In terms of the function $h$ we are trying to solve for,
$\Theta_{(\ell)}$ depends on $h$ and its first two derivatives.  We
thus end up with a second order non-linear elliptic partial differential
equation for $h$ whose coefficients depend on $h_{ij}$, its first
derivatives, and $K_{ij}$.

The method using the above formalism, with the assumption of a
star-shaped surface, i.e. using Eq.~\eqref{eq:starshaped}, and
following the implementation of \cite{Thornburg:2003sf} is routinely
used in numerical simulations.  While this is generally sufficient for
many applications, there are cases where highly distorted MOTSs
appear. An example is in \cite{Gupta:2018znn} where certain highly
distorted MOTSs appear and this standard approach does not work.  We
will see that this restriction can be removed by a small change while
still leaving most aspects of the above approach intact.

\section{Locating a distorted MOTS}
\label{sec:motsfinder}

We begin by reinterpreting the starting point of the algorithm,
i.e. Eq.~\eqref{eq:starshaped}.  Instead of the radial distance $r$
from the origin, we could choose to use the distance from a sphere of
some radius $R_0$ along rays orthogonal to the sphere\footnote{Here
  orthogonality refers just to a Euclidean metric in the coordinate
  system where the numerical calculations are being performed.
  Similarly straight lines refer to this fictitious Euclidean metric
  and not to geodesics of the physical Riemannian metric.}  (the
precise radius of the sphere is not important).  Since the rays
emanating from the sphere in the orthogonal direction all meet at the
origin, the two interpretations are identical and the numerical method
and results are unchanged. The restriction to a star-shaped surface
obviously still holds.  However, we are free to take as reference a
topologically spherical surface of any arbitrary shape and we can
consider rays orthogonal to it.  The rays will now not necessarily
meet at the origin (or at any other point), but this is irrelevant.
What is important is that the reference surface can be chosen so that
the rays orthogonal to it meet the surface $\mathcal{S}$ just once.
The reference surface itself need not be star-shaped either, it just
needs to be parameterized suitably, as we shall discuss later.
Furthermore, it is important that the rays do not intersect each other
\emph{before} reaching $\mathcal{S}$.  This typically happens when
$\mathcal{S}$ is too far away from the reference surface; it is beyond
the region of validity of the coordinate system based on the reference
surface.

\begin{figure}
  \centering
  \includegraphics[width = 0.45\textwidth]{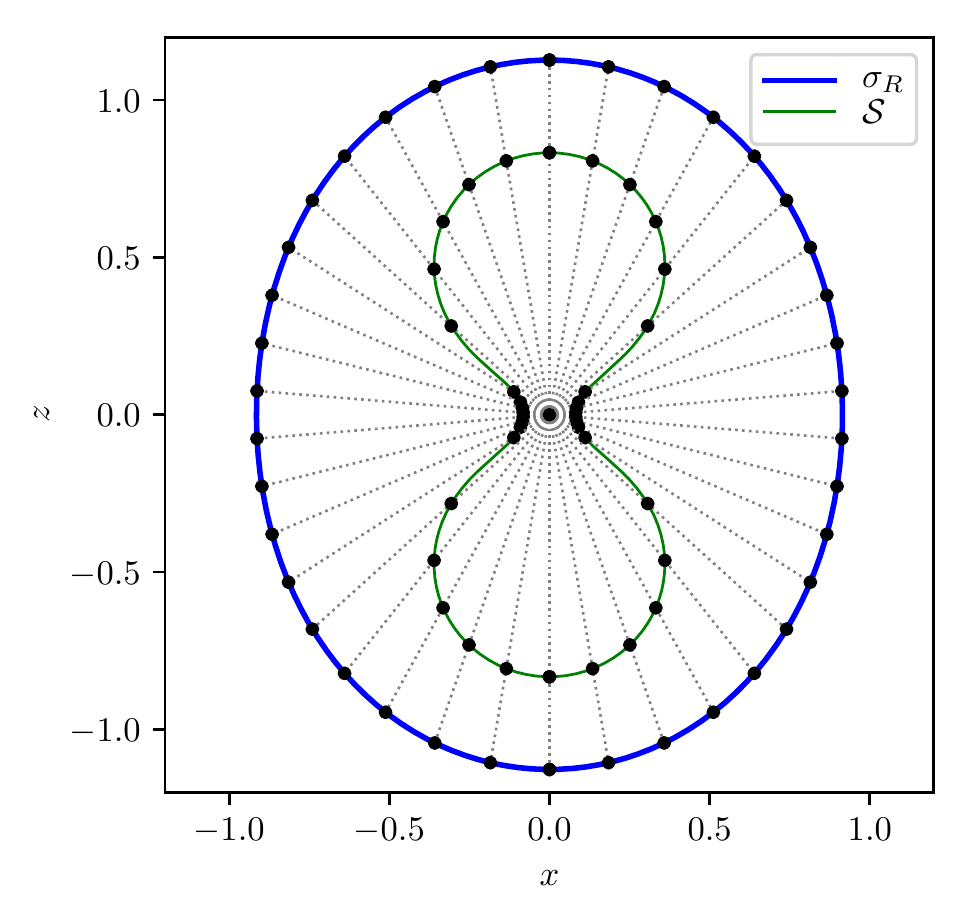}
  \includegraphics[width = 0.45\textwidth]{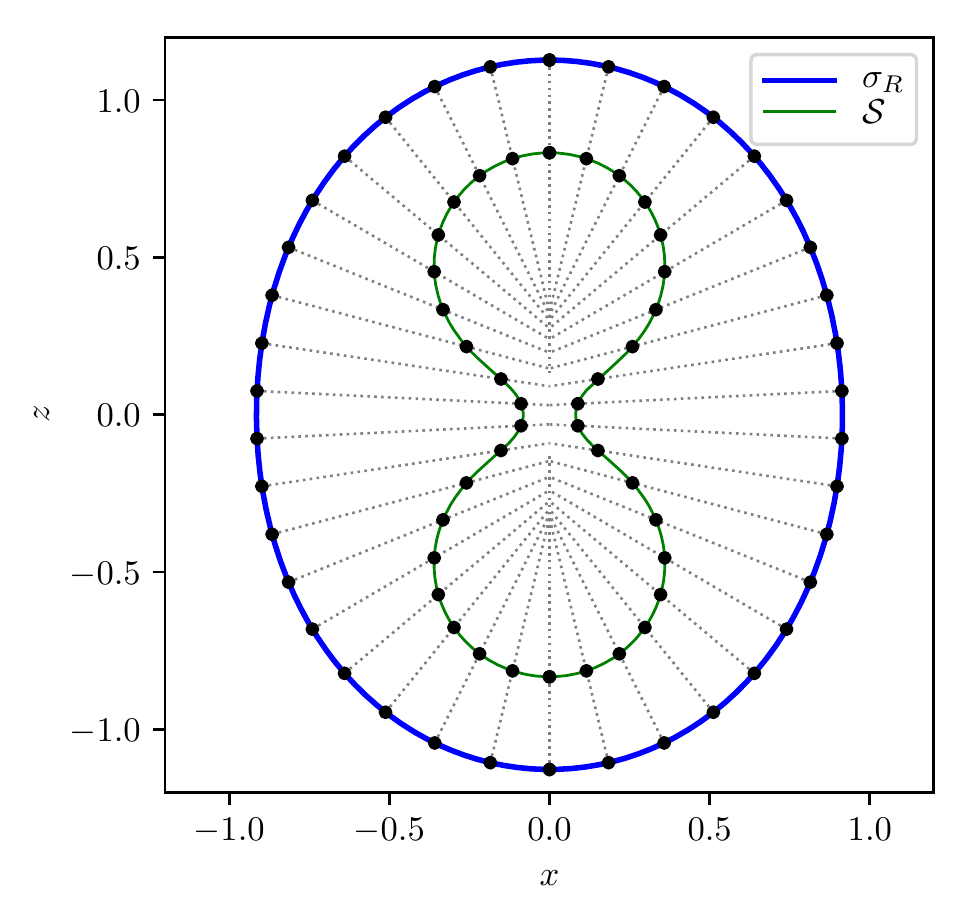}
  \caption{This figure shows the benefits of using a non-spherical
    reference surface.  The surface $\mathcal{S}$ in both panels is
    the MOTS we are trying to locate; it is star shaped, but only
    barely so.  The first panel is the standard approach using rays
    centered at the origin while the second panel uses rays orthogonal
    to a reference surface $\sigma_R$.  See text for further
    discussion.}
  \label{fig:reference_surface_example}
\end{figure}

An example of the numerical benefits of choosing a suitable reference
surface is shown in Fig.~\ref{fig:reference_surface_example}.  The
surface $\mathcal{S}$ is the MOTS that we are trying to locate.  The
first panel of Fig.~\ref{fig:reference_surface_example} is the
standard method, i.e. using rays centered at the horizon.  The surface
$\mathcal{S}$ is of the type that is difficult to locate and comes
close to, or even violates, the property of being star shaped; this is
very similar to Fig. 1.1 of \cite{Thornburg:2006zb} shown there as an
example of a problematic surface.  In
Fig.~\ref{fig:reference_surface_example}, the surface $\mathcal{S}$ is
actually star shaped, but only barely.  We see that the rays centered
at the origin intersect $\mathcal{S}$ only once, but the intersections
pile up at the neck.  On the other hand, the second panel shows a
reference surface $\sigma_R$ which is only slightly deformed away from
an exact sphere.  The rays are now orthogonal to $\sigma_R$ and we see
that the intersections are much more uniformly distributed on
$\mathcal{S}$.  While it is clearly possible to find reference
surfaces which make it harder to locate $\mathcal{S}$, there are many
easily found choices which greatly improve the numerical results. This
also makes the algorithm very flexible, as it can be tuned to find
extremely distorted surfaces.

\subsection{The new coordinate system}
\label{subsec:parameterization}

\begin{figure}
  \centering
  \includegraphics[scale=0.4]{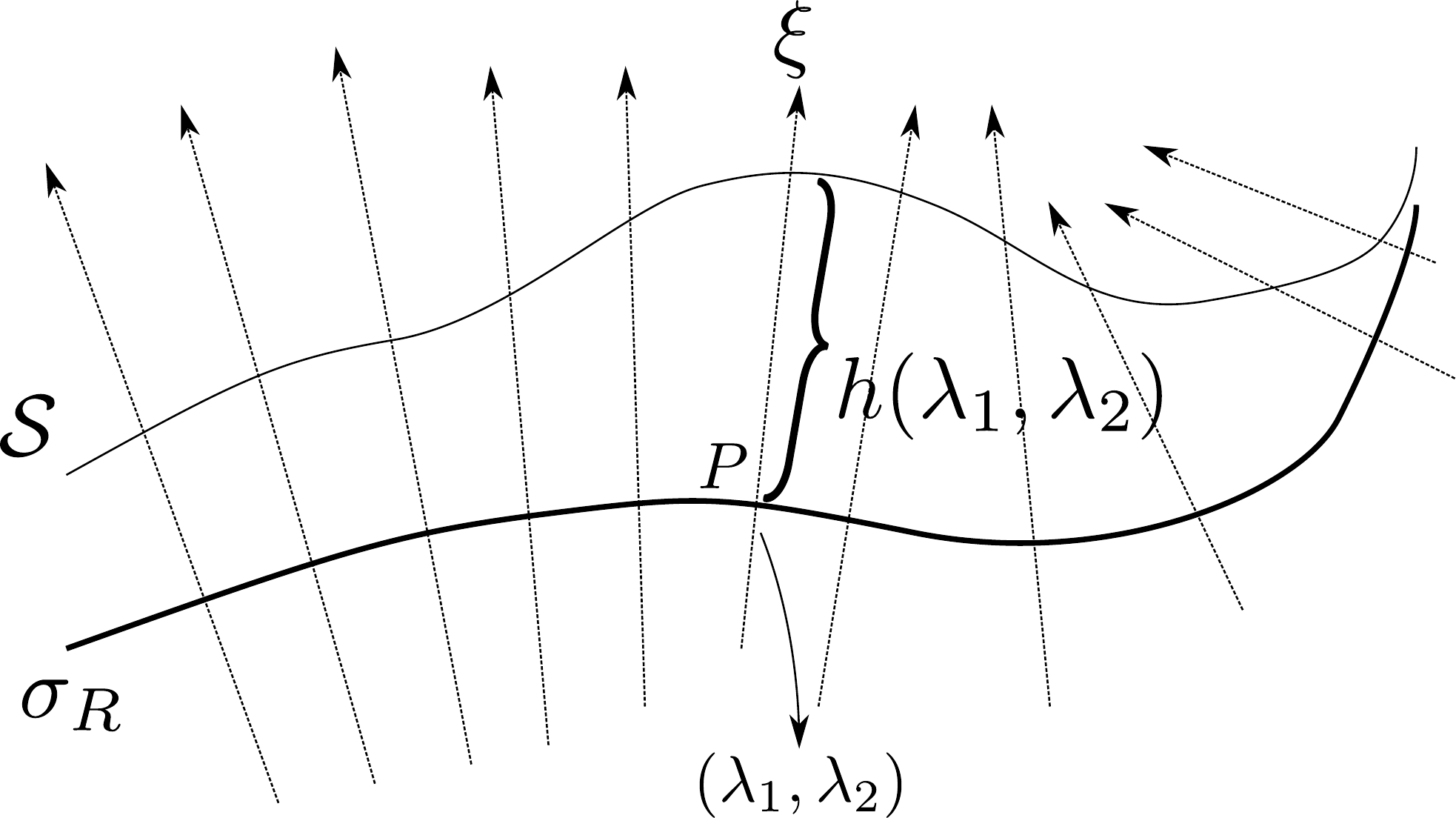}
  \caption{The coordinate system $(\xi,\lambda_1,\lambda_2)$ based on a
    reference surface $\sigma_R$ and the rays orthogonal to it.  $\xi$
    is the Euclidean distance along the orthogonal rays and
    $(\lambda_1,\lambda_2)$ are the coordinates of the point $P$ which
    is the intersection of the ray and $\sigma_R$.  The surface
    $\mathcal{S}$ we are looking for can be represented by a height
    function $h$.}
  \label{fig:reference_surface}
\end{figure}

Consider then a reference surface $\sigma_R$ parameterized by two
coordinates $(\lambda_1,\lambda_2)$ (the generalization to greater or
fewer dimensions is obvious).  The parameters $(\lambda_1,\lambda_2)$
could be for example angular coordinates $(\theta,\phi)$ but this is
not a requirement. Construct then the rays orthogonal to $\sigma_R$
using the Euclidean metric in which the numerical simulation is being
carried out.  To any point in the neighborhood of $\sigma_R$, we can
assign coordinates $(\xi,\lambda_1,\lambda_2)$ where $\xi$ is the Euclidean
distance along the orthogonal rays; see
Fig.~\ref{fig:reference_surface}.  This coordinate system is valid as
long as the orthogonal rays do not cross.  This construction is very
similar to Gaussian or Fermi-normal coordinates in differential
geometry, except that we do not use the actual curved geometry to
define the orthogonal rays, nor do we use the proper distance
along the rays.

Given the coordinate system $(\xi,\lambda_1,\lambda_2)$, we can
represent the surface $\mathcal{S}$ that we are looking for by a
height function $h$.  Then, analogous to Eq.~\eqref{eq:starshaped}, we
want to consider level sets of the function
\begin{equation}\label{eq:nonstarshaped}
  F(\xi,\lambda_1,\lambda_2) = \xi - h(\lambda_1,\lambda_2)\,.
\end{equation}
The normal to $\mathcal{S}$ is again the gradient of $F$:
$s_i = \partial_iF$.  Obviously, $(\xi,\lambda_1,\lambda_2)$ must be
known as functions of the coordinates used in the simulation.  The
important point is that Eq.~\eqref{eq:expansion_AB} still remains
valid.  The difference now is that instead of
Eq.~\eqref{eq:starshaped}, we use Eq.~\eqref{eq:nonstarshaped} to
define the normal $s_i$.  The numerical method for solving the
equation is not affected by this parameterization.

While it should be clear that the method should work generally, in the
rest of this paper, we shall restrict ourselves to the axisymmetric
case where we only need to consider reference curves parameterized by a
single parameter $\lambda$.  The validation of the code and results
for the general 3-dimensional case will be left to future work.  Let
the symmetry axis be the $z$-axis, and let us work in the $x-z$ plane.
We represent a reference surface $\sigma_R$ via a parameterized curve
which we shall denote $\gamma_R(\lambda)$.  We are free to choose the
parameter $\lambda$ as convenient.  For example we could take it to be
the angle with the $z$-axis - but this would restrict us to take the
reference surface as star-shaped. More generally, we could take it to
be the path-length of the curve.  Whatever the choice, we shall take
the range of the parameters to be from $0$ to $\pi$.  Thus we have a
curve $\gamma_R : (0,\pi) \to \mathbb{R}^2$ in the $x$-$z$-coordinate
plane such that $\gamma_R(0)$ and $\gamma_R(\pi)$ lie on the $z$-axis
and the tangent vectors $\gamma_R'(0)$ and $\gamma_R'(\pi)$ are
perpendicular to the $z$-axis.

Let $\vec{x}$ be the vector representing any point in the $x-z$ plane;
we remind the reader again that the simulations are in Euclidean
coordinates. We define
\begin{equation}\label{eq:F_non_star_shaped}
  F(\vec{x}) := \xi(\vec{x}) - h(\lambda(\vec{x})),
\end{equation}
where $\xi(\vec{x})$ and $\lambda(\vec{x})$ are defined implicitly
by
\begin{equation}\label{eq:d_l_inplicit}
    \vec{x}(\xi,\lambda) = \vec{\gamma}(\lambda) + \xi\;\vec{\nu}_R(\lambda).
\end{equation}
Here $\vec{\nu}_R$ is a vector pointing outwards in the direction normal
to $\sigma_R$ in the (Euclidean) $x,z$ coordinates.  Note that we do not require
$\vec{\nu}_R$ to be normalized to unit length.  This will make the
computational tasks much easier, since it allows us to simply choose
$\nu_R$ to be the tangent vector rotated by $\pi/2$.

Using this ansatz, the horizon function $h$ defines the surface
$\mathcal{S}$ (for which $F=0$) via the curve
\begin{equation}\label{eq:sigma_curve}
  \vec{\gamma}(\lambda) = \vec{\gamma}_R(\lambda) + h(\lambda) \vec{\nu}_R(\lambda)\,.
\end{equation}
As before, taking the normal $s_i$ to this curve and using it in
Eq.~\eqref{eq:expansion_AB} yields the differential equation that we
need to solve.

\subsection{The numerical algorithm}
\label{subsec:algorithm}

As discussed above, the equation to solve numerically is
\eqref{eq:expansion_AB} with our definition of $F$ from
Eq.~\eqref{eq:F_non_star_shaped} used to define the normal $s_i$.
This is then read as a non-linear partial differential equation for $h$,
which becomes a non-linear ordinary differential equation in
axisymmetry.
Our method of solving this equation is
standard Newton root-finding, suitably extended to differential
operators via the Newton-Kantorovich scheme \cite{BoydBook}.  Other
means of solving the non-linear PDE, e.g. \cite{Thornburg:2006zb},
can also be used.  Let $\mathcal{N}$ be the differential operator so
that the equation to be solved is $\mathcal{N}(u) = 0$.  We will need
the variational derivative $\mathcal{N}_u$ of $\mathcal{N}(u)$ defined
in the usual manner:
\begin{equation}\label{eq:linearized_op_definition}
  \mathcal{N}_u(\Delta) = \lim_{\epsilon\rightarrow 0} \frac{\mathcal{N}(u+\epsilon\Delta) - \mathcal{N}(u)}{\epsilon}
\end{equation}
Here $\Delta$ is a variation of $u$ and $\mathcal{N}_u$ is a
differential operator (obtained by linearizing $\mathcal{N}$) acting
on $\Delta$.

Start with an initial guess $u^{(0)}$, and let $u^{(i)}$ be the
$i^{\rm th}$ iteration.  If this is sufficiently close to the true
solution, we can expand $\mathcal{N}(u)$ linearly around $u^{(i)}$
to obtain
\begin{equation}
  0 = \mathcal{N}(u^{(i)}+\Delta) \approx \mathcal{N}(u^{(i)}) + \mathcal{N}_u\Delta \,.
\end{equation}
We want to choose $u^{(i+1)}$ by solving the linear system of equations
\begin{equation}
  \label{eq:newton_step}
  u^{(i+1)}= u^{(i)} - \mathcal{N}_u^{-1}\mathcal{N}(u^{(i)}),
\end{equation}
i.e. we solve $\mathcal{N}_u \Delta = -\mathcal{N}(u^{(i)})$ for
$\Delta$ (with suitable boundary conditions to be detailed below) and
perform the steps $u^{(i+1)} = u^{(i)} + \Delta$.
This applies whether $u$ is represented by values on a discrete grid,
or as spectral coefficients.

Our implementation uses a pseudospectral method \cite{BoydBook,Canuto2006Book} to
perform the steps \eqref{eq:newton_step}, where
$u^{(i)}(\lambda) = h^{(i)}(\lambda)$ and $\mathcal{N}(u) = \Theta(h)$
as defined in \eqref{eq:expansion_AB}.
The linear equation for $\Delta$ now takes the form
\begin{equation}\label{eq:linear_eq_for_Delta}
    (\delta_h\Theta) \Delta
    + (\delta_{h'}\Theta) \Delta'
    + (\delta_{h''}\Theta) \Delta''
        = -\Theta(h^{(i)}).
\end{equation}
Here, the derivatives of $\Theta$ on the left-hand side of
\eqref{eq:linear_eq_for_Delta} are evaluated at $h^{(i)}$.
Note that in each step, $h^{(i)}$ defines a {\em trial surface}
$\mathcal{S}^{(i)}$ which is tested for convergence by computing its
expansion $\Theta$ at a finite set of points.

To apply the pseudospectral method, we start with a choice for
$h^{(0)}$, say $h^{(0)} \equiv 0$, and represent $\Delta$ using a truncated
series of cosines,
\begin{equation}\label{eq:h_cos_series}
    \Delta(\lambda) = \sum_{n=0}^N a_n \cos(n\lambda),
\end{equation}
where we call $N$ the {\em resolution} of a particular pseudospectral representation.
This choice of basis functions is natural for the axisymmetric setting
and ensures that for a reference curve $\gamma_R$ satisfying
$\gamma'_R(0) = \gamma'_R(\pi) = 0$, the curve represented by $h$ using
\eqref{eq:sigma_curve} does so too. In addition, it eliminates the need
to impose explicit boundary conditions when solving the differential
equation since this is already done by each individual basis function.

\subsection{Validating the numerical method}
\label{subsec:convergence}

\begin{figure}
  \centering
  \includegraphics[trim=0 5 0 0,clip,width=1.0\linewidth]{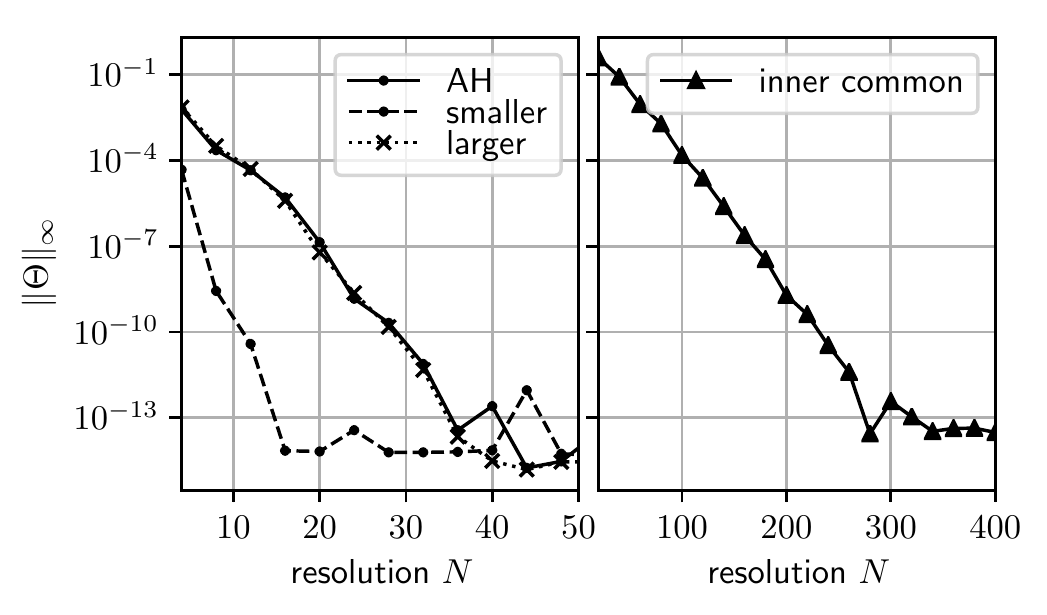}
  \caption{%
    Convergence of the expansion using results of searches with different spectral
    resolutions $N$ for the series representation of $h^{(i)}$.
    The quantity $\Vert\Theta\Vert_\infty$ is computed taking an initial
    grid with higher density than the resolution $N$ and performing a
    local maximum search from the point of largest deviation from zero.
    The left panel shows the convergence of the outermost common (apparent) horizon and
    the two individual MOTSs, while the right panel shows the convergence for the inner
    common MOTS which has a more distorted shape and hence requires a higher
    resolution.
    The expected exponential fall-off continues until the floating point
    roundoff plateau is reached.
    As can be seen, this happens at different resolutions depending on the
    specific form of the respective surface $\mathcal{S}$ and the reference
    surface $\sigma_R$.
    The configuration used here is that of Fig.~\ref{fig:sub3} with reference
    surfaces taken as the respective horizons for $d=0.405$.
  }
  \label{fig:convergence_Ta_of_N}
\end{figure}

Since the full procedure combines several methods (computation of the
variational derivatives in \eqref{eq:linear_eq_for_Delta},
the Newton-like root search, the pseudospectral method), each introducing
numerical errors, one possibility for testing the procedure as a whole is to perform
the whole search for $\mathcal{S}$ at different fixed spectral resolutions $N$
of the steps $h^{(i)}$.
At each resolution, we compute the maximum $\Vert \Theta \Vert_\infty$ and
plot the result as a function of $N$ in Fig.~\ref{fig:convergence_Ta_of_N},
for a static initial data configuration of two black holes (as described in
the next section \ref{sec:brill}).
The exponential convergence is expected for a spectral method \cite{BoydBook} and also shows
convergence of the Newton-Kantorovich scheme until the floating point roundoff
plateau is reached.

\begin{figure}
  \centering
  \includegraphics[width=0.4\textwidth]{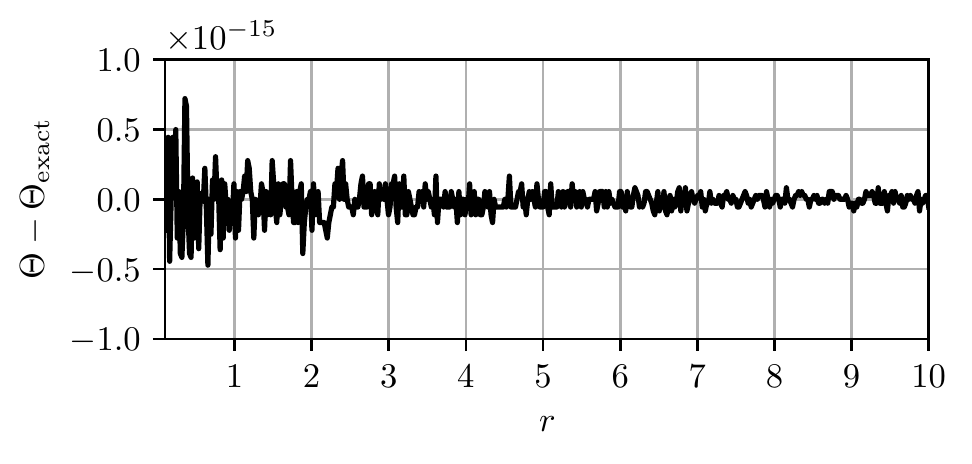}
  \caption{%
    Accuracy of the expansion computed for an $r = \mathrm{const}$ surface in
    a Schwarzschild slice with mass $m = 1$. The reference surface $\sigma_R$
    is here chosen to be a sphere of radius $r/2$.
  }
  \label{fig:schwarzschild_expansion}
\end{figure}

The remaining quantity to be tested is the expansion as computed for a (trial)
surface $\mathcal{S}$. This can be accomplished by comparing the computed
expansion with a case where it is known analytically.
One of the simplest of such cases is a centered sphere in a Schwarzschild
slice $h_{ab} = \phi^4 \delta_{ab}$, where
\begin{equation}\label{eq:schwarzschild_conformal}
    \phi(\vec{r}) = \phi(r) = 1 + \frac{m}{2r}.
\end{equation}
The expansion of $r = \mathrm{const}$ surfaces is then given by
\begin{equation}\label{eq:schwarzschild_expansion}
    \Theta = \frac{2}{\phi^2(r)} \left( \frac{1}{r} + 2 \frac{\phi'(r)}{\phi(r)} \right).
\end{equation}
Fig.~\ref{fig:schwarzschild_expansion} shows that the numerically
computed expansion agrees to high accuracy with the exact value.  The
event horizon and apparent horizon coincide in these slices and are
located at $r = m/2$ in these coordinates. A respective search with a
reference surface chosen to be $r = m = 1$ converges to a curve
$\gamma$ with $|\gamma(\lambda)| = m/2 \pm 5 \times 10^{-17}$ and a
numerically evaluated expansion
$\Vert\Theta\Vert_\infty \lesssim 10^{-15}$.

\section{Application to time symmetric binary black hole initial data}
\label{sec:brill}

\subsection{Brill-Lindquist initial data}

\begin{figure*}
  \begin{subfigure}{0.48\textwidth}
    \centering
    \includegraphics[width=\textwidth]{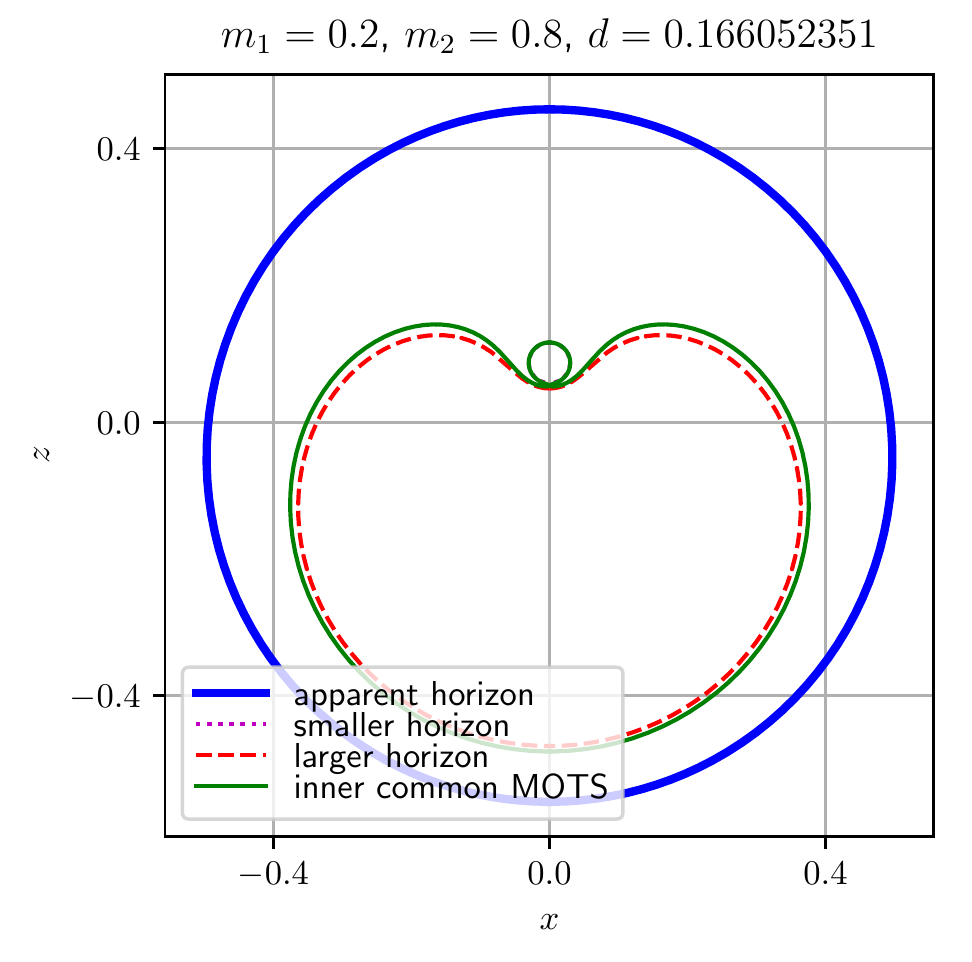}
    \caption{The marginal surfaces just before the extremely distorted inner common MOTS
      disappears entirely}
    \label{fig:sub5}
    \centering
  \end{subfigure}
  \hfill
  \begin{subfigure}{0.48\textwidth}
    \centering
    \includegraphics[width=\textwidth]{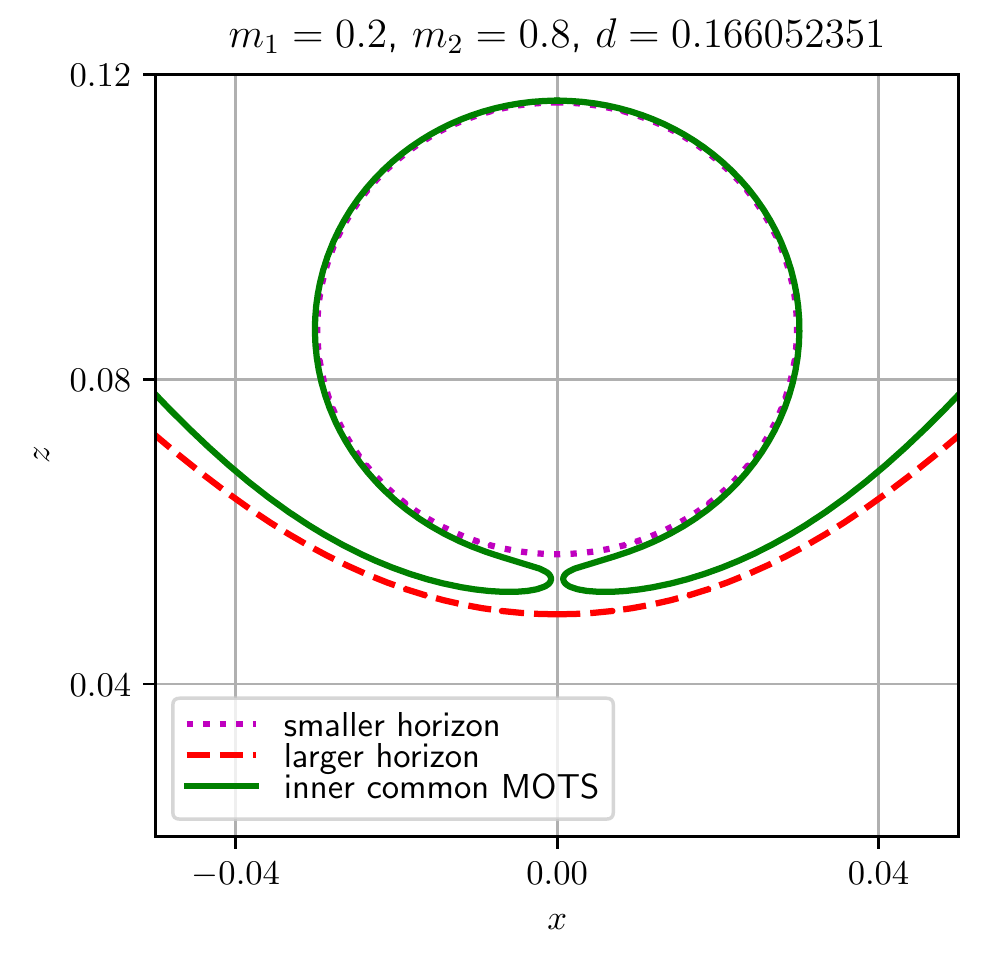}
    \caption{Detailed view of the inner MOTS at its neck
	- where it almost seems to be pinching off -
	just before it disappears.}
    \label{fig:sub6}
    \centering
  \end{subfigure}
  \caption{Same as Fig. \ref{fig:brill_lindquist1} above, but
    for the smallest value of $d$ before the inner common MOTS
    disappears.  The right panel \ref{fig:sub6} shows the details of
    the highly distorted inner MOTS at its neck. }
  \label{fig:brill_lindquist2}
\end{figure*}

We apply our new MOTS finder algorithm to the
Brill-Lindquist (BL) initial data set.  This is perhaps the simplest
initial data set \cite{Brill:1963yv} representing multiple
non-spinning black holes at a moment of time symmetry.  Thus, each
black hole has vanishing momentum.  The original work
\cite{Brill:1963yv} considered also electric charge and an arbitrary
number of black holes, but here we shall ignore charge and restrict
ourselves to two black holes.  While simple, this initial data set is
not simplified in terms of the various kinds of MOTSs that can appear.
As shown numerous times in the literature, see
e.g. \cite{Gupta:2018znn}, the general picture is that when the two
black holes in the binary are initially well separated, there are two
independent MOTSs, one for each black hole. These two MOTSs approach each
other and at a certain point, a common MOTS appears which surrounds
the individual MOTSs.  This common MOTS immediately bifurcates into an
inner and outer MOTS.  The inner common MOTS shrinks and approaches
the two individual MOTSs, while the outer common MOTS (the apparent
horizon) grows and sheds its multipole moments to approach an
equilibrium state, i.e. a Kerr black hole\footnote{or in our simplified
case of axial symmetry and no spins, a Schwarzschild black hole.}.
The eventual fate of the
inner MOTS and the two individual MOTSs is still unknown, though some
partial results are known \cite{Gupta:2018znn,Mosta:2015sga}.  

Time symmetry means that the extrinsic curvature vanishes:
$K_{ab} = 0$.  The 3-metric is conformally flat:
$h_{ab} = \phi^4\delta_{ab}$.  The two black holes are represented by
two punctures in the background conformal metric $\delta_{ab}$.  The
Euclidean distance between the punctures is $d$ and the conformal
factor at any point $\vec{r}$ is
\begin{equation}
  \label{eq:conformal_factor}
  \phi(\vec{r})  = 1 + \frac{m_1}{2r_1} + \frac{m_2}{2r_2} \,.  
\end{equation}
Here $r_1$ and $r_2$ are the Euclidean distances of $\vec{r}$ from the
two punctures and $m_1,m_2$ are the bare masses of the two black
holes.  As shown in \cite{Brill:1963yv}, the total ADM mass of the
system is $M_{ADM} = m_1+m_2$.  The two punctures are regular
asymptotic ends, and ADM masses can thus be calculated at the
punctures:
\begin{eqnarray}
  M^{(1)}_{ADM} &=& m_1 + \frac{m_1m_2}{2d}\,,\\
  M^{(2)}_{ADM} &=& m_2 + \frac{m_1m_2}{2d}\,.
\end{eqnarray}
The difference
\begin{equation}
  M_{ADM} - M^{(1)}_{ADM} - M^{(2)}_{ADM} = -\frac{m_1m_2}{d}
\end{equation}
is interpreted as the binding energy. For non-spinning black holes,
the irreducible mass, i.e. $M_{irr} = \sqrt{A/16\pi}$ with $A$ being
the area of the MOTS, provides an appropriate notion of horizon mass.
The difference between the bare masses and the horizon masses
($M_{ADM} - M^{(1)}_{irr} - M^{(2)}_{irr}$) is interpreted similarly,
yields similar results, is applicable for a much wider variety of
initial data, and is in fact physically more meaningful
\cite{Schnetter:2006yt,Krishnan:2002wxg}.

The qualitative structure of the MOTSs described above as arising from
time evolution is also present when the static distance $d$ is
reduced. As far as the various MOTSs are concerned, there is no
reduction in complexity by simplifying the initial data.  The main
difference is that for time symmetric data a MOTS is also a surface of
extremal area.  As in time evolution, and as shall be shown below, all
of the individual and the inner and outer MOTSs are present.  As $d$
becomes smaller, the inner common MOTS becomes highly distorted and
extant horizon finders have not been able to find it.  With our new
numerical method, we shall be able to track this inner MOTS even as it
becomes highly distorted and eventually disappears.  It should be kept
in mind however that changing $d$ alone is not equivalent to time
evolution and quantitative results may not carry through to a real
time evolution.  In a real time evolution the data will not remain
time symmetric.  Also, decreasing $d$ alone keeps the total ADM mass
fixed but leads to increasing $M^{(1,2)}_{ADM}$ as is obvious from the
above equations. A true time evolution will, in principle, keep all
ADM masses fixed.

\subsection{Results: Distorted MOTSs in BL data}
\label{subsec:blresults}

We can now proceed with presenting our main results, namely properties
of marginal surfaces in BL data for various values of the masses
$m_{1,2}$ and separations $d$.  By convention, we shall take
$m_1\leq m_2$ and the mass ratio is defined as $q=m_2/m_1 \geq 1$. The
total mass will be kept fixed to unity: $M=m_1+m_2 = 1$.  For any
given mass ratio, we generally successively decrease the value of $d$
in the results below.  Once we have found a MOTS for a particular $d$,
it is used as a reference surface for the next smaller value of $d$.

\subsubsection{The various MOTSs for mass ratio 1:4}
\label{subsec:massratios}

The first set of plots shows the various MOTSs for BL data with a mass
ratio 1:4, chosen as an illustrative example.  This value of the mass
ratio is also the one used in \cite{Mosta:2015sga} as an example, and
we shall reproduce and extend the results shown in Fig.~2 of
\cite{Mosta:2015sga}.

Since the total mass is normalized to unity, the individual masses are
$m_1 = 0.2$, $m_2= 0.8$. Note that since the data is time symmetric,
the MOTSs are all minimal surfaces and thus they cannot touch each
other with a common tangent vector (otherwise, by the maximum
principle for elliptic operators, they must coincide
\cite{Mosta:2015sga}).  The common horizon is seen to exist for
$d\leq 0.6987162$.\footnote{%
    We remark that for the equal mass case we find the common horizon
    for $d \leq 0.76619742$, which agrees within our final step size
    $\Delta d = 10^{-8}$ with the results found in
    \cite{Lages2010Dissertation} by which the horizon should cease to
    exist one step further out.
  }
For larger separations we only have the two individual horizons.
Figs.~\ref{fig:brill_lindquist1} and ~\ref{fig:brill_lindquist2} show
the horizons for $d=0.69850,0.6500, 0.4000, 0.2500, 0.166052351$.
This last value is just before the inner common MOTS ceases to exist
and the last panel, Fig.~\ref{fig:sub6}, shows the detailed picture
near the neck of the MOTS. It is clear that the inner common MOTS is
far from star-shaped at this point (in fact much earlier for larger
$d$), yet the new horizon finder has no fundamental difficulty in
locating it.

For each of the MOTSs, we also look for surfaces of constant expansion
\cite{Schnetter:2003pv,Schnetter:2004mc}
on both sides.  We confirm that indeed the behavior of the apparent
horizon is as expected, i.e. the expansion goes smoothly from negative
to positive values as we cross the apparent horizon going outwards.  
This is \emph{not} the case for the inner MOTS.  The constant
expansion surfaces lie on both sides of the MOTS and thus they do not
form a regular foliation in any neighborhood of the MOTS; see
Fig.~\ref{fig:CE}. This is consistent with the fact that, as we shall
see below, the inner MOTS (unlike the apparent horizon) is unstable in
the sense discussed in Sec.~\ref{subsec:definitions} and it
thus is not guaranteed to be a barrier for trapped and untrapped surfaces
in any neighborhood \cite{Andersson:2005gq}.


%
\begin{figure*}[h]
    \includegraphics[width=0.90\textwidth]{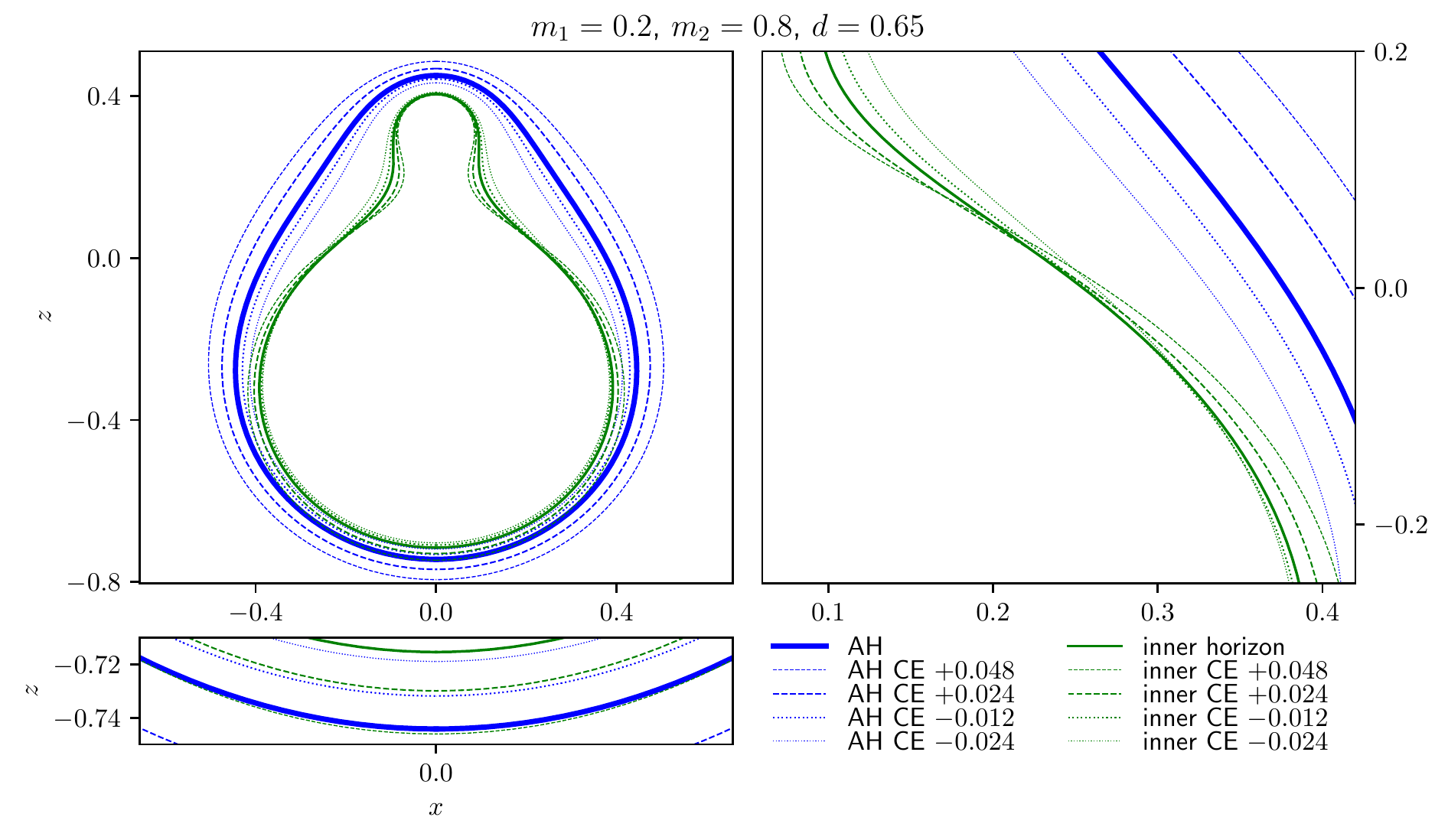}
    \caption{The constant expansion surfaces near the apparent horizon
      and the inner common MOTS.  This shows that the apparent horizon is
      stable while the inner MOTS is not.  See text for discussion.
      The right panel shows a close-up of the intersection of surfaces
      of constant positive and negative expansion with the inner MOTS.
      The bottom panel shows that the constant expansion surfaces
      and MOTSs do not coincide and even though being close, they have
      different curvatures leading to different expansion values.}
    \label{fig:CE}
\end{figure*}
%


As we decrease $d$, we observe that the inner common MOTS is not found
for $d< 0.166052351$, and shortly after that, for $d< 0.16461385$ the
larger individual MOTS is not found as well.  For all smaller values
of $d$, the smaller individual horizon and the apparent horizon continue to
exist; the apparent horizon becomes smoother just as in a time
evolution \cite{Gupta:2018znn}.  The remaining inner MOTS
corresponding to the smaller black hole becomes smaller in coordinate
space.  In previous work, say \cite{Gupta:2018znn}, when the inner
horizon was not found, it was reasonable to suspect that the numerical
method being used there was not able to find it.  Here, we have
evidence that the inner horizon in fact does not exist.  While it is
impossible to prove non-existence with absolute certainty here, we can
eliminate a few clear possibilities.

First, we note that at the point before the MOTS disappears, there is
no indication of any numerical problems at the earlier step (in this
case for a slightly larger value of $d$).  It is true that resolving
the neck requires higher resolution, but this is feasible with modern
computers.  Note also that the larger individual MOTS disappears as
well and it does not have any such features that need to be resolved.
Second, the foliation by the constant expansion surfaces, in the
region near where the inner MOTS is found just before it disappears,
shows only negative values of the expansion: $\Theta_{(\ell)} < c$
where $c$ is some non-zero negative number.  Just before the MOTS
disappears, $c$ is positive and decreases to $0$ when the last
instance of the MOTS is found. This shall be further discussed below
in more detail and for other mass ratios.

\subsubsection{Varying the mass-ratio}
\label{subsubsec:massratios}

We next investigate a number of geometric quantities on the various
MOTSs as functions of $d$, and for different mass ratios.  In these
results, we consider the mass ratios 1:1, 1:2, 1:3, 1:4 and 1:5.  As
before the total mass is always kept fixed: $M_{ADM} = m_1+m_2=1$.

\begin{figure*}[htpb]
  \begin{subfigure}{.45\textwidth}
    \centering
    \includegraphics[width=\textwidth]{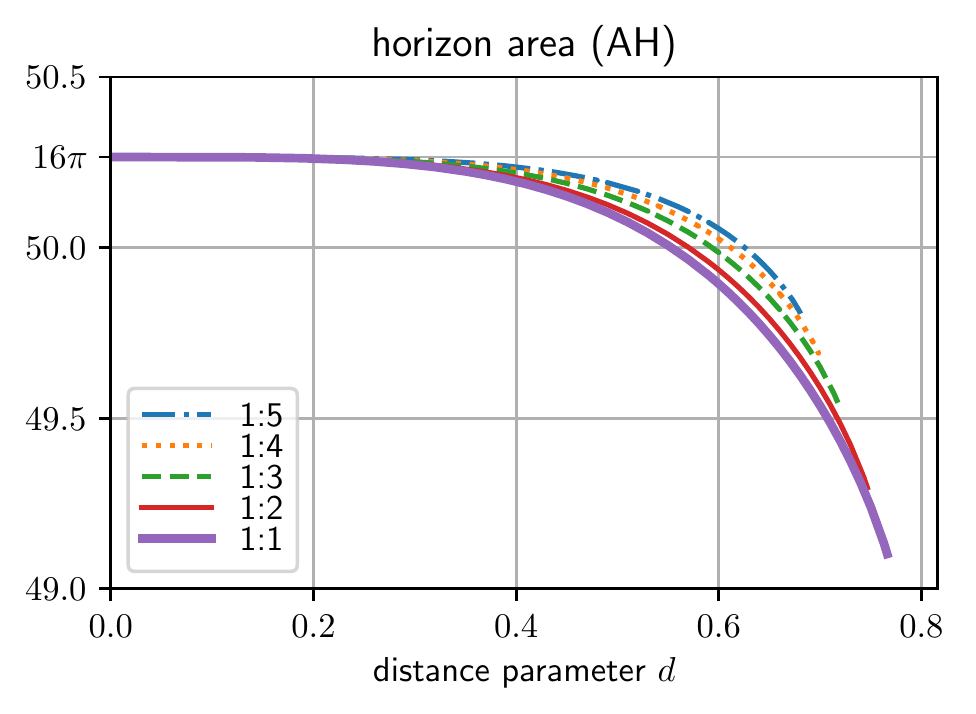}
    \caption{The area of the AH as a function of $d$ for different
      mass ratios. As discussed in the text, we get $16\pi$ as
      $d\rightarrow 0$. }
    \label{fig:sub_area1}
  \end{subfigure}
  \hfill
  \begin{subfigure}{.45\textwidth}
    \centering
    \includegraphics[width=\textwidth]{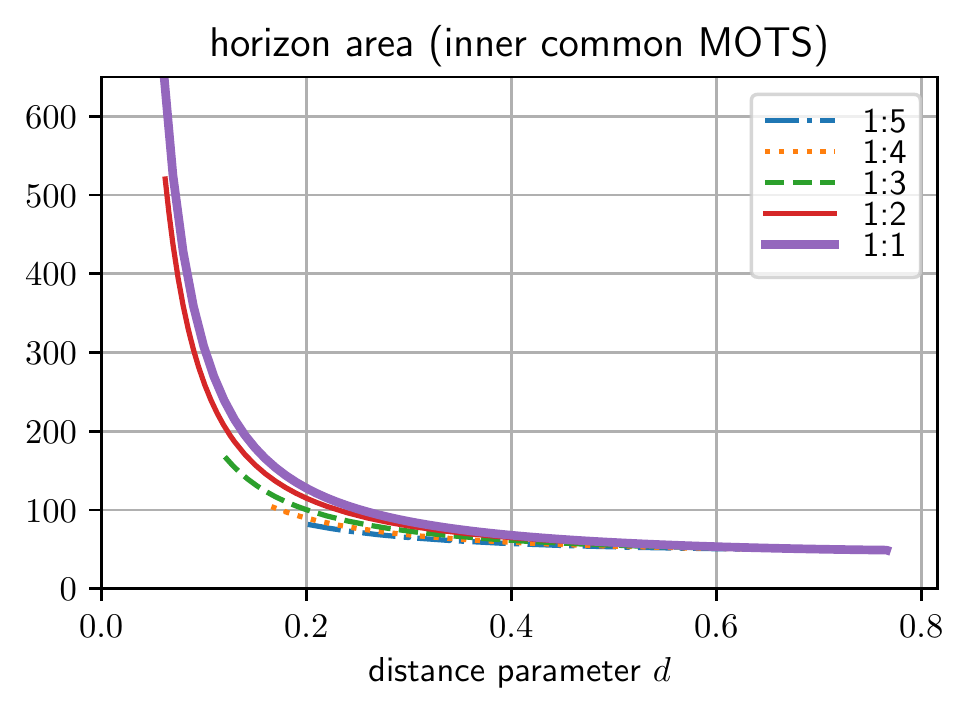}
    \caption{The area of the inner MOTS as a function of $d$ for
      different mass ratios, for all values of $d$ for which the MOTS
      exists.  For smaller mass ratios the MOTS continues existing for
      smaller $d$. }
    \label{fig:sub_area2}
  \end{subfigure}
  \hfill  
  \begin{subfigure}{.45\textwidth}
    \centering
    \includegraphics[width=\textwidth]{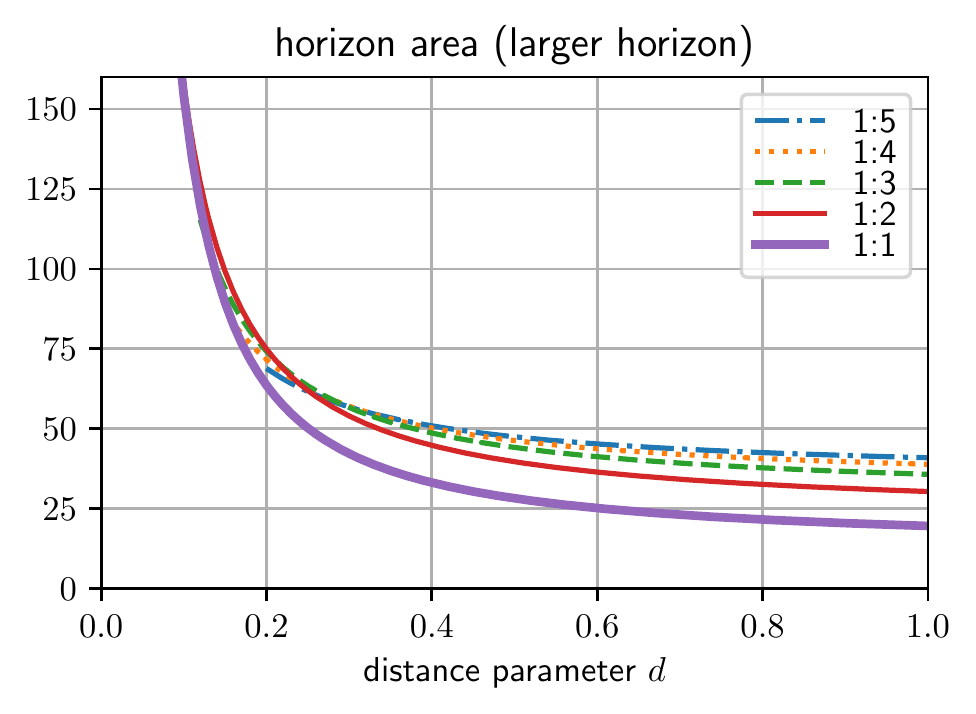}
    \caption{The area of the larger individual MOTS as a function of
      $d$ for different mass ratios. As shown in the text, for large
      $d$, the area approaches $16\pi q^2/(1+q)^2$, i.e. becomes larger
      for more asymmetric systems.}
    \label{fig:sub_area3}
  \end{subfigure}
  \hfill
  \begin{subfigure}{.45\textwidth}
    \centering
    \includegraphics[width=\textwidth]{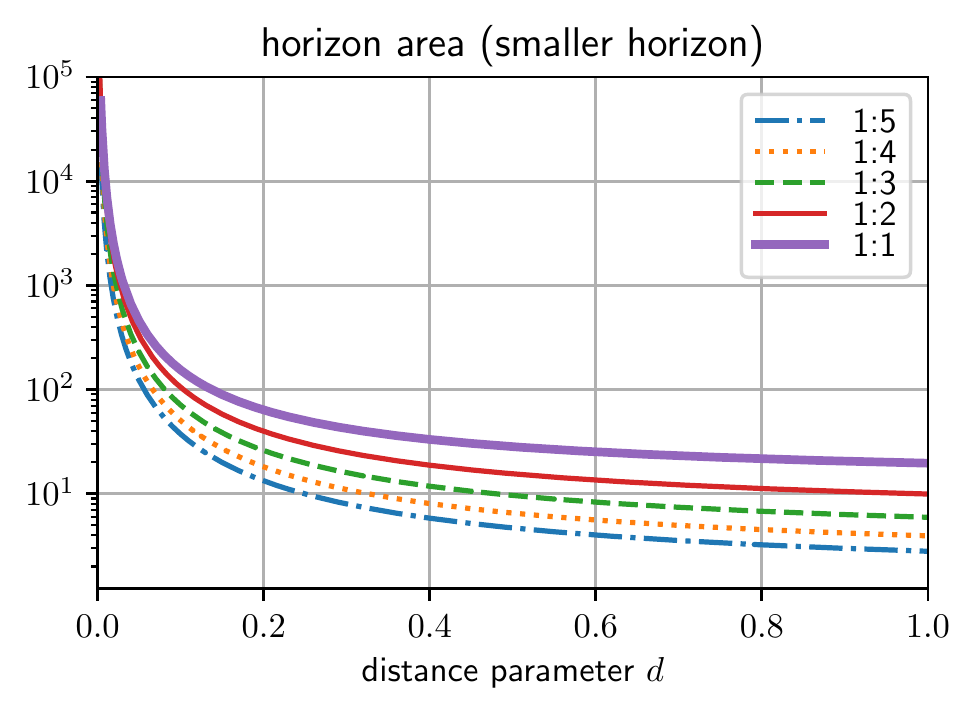}
    \caption{The area of the smaller individual MOTS as a function of
      $d$ for different mass ratios. As shown in the text, for large
      $d$, the area approaches $16\pi/(1+q)^2$, i.e. becomes smaller for
      more asymmetric systems.}
    \label{fig:sub_area4}
  \end{subfigure}
  
  \caption{The area of the various horizons in BL data
    with different mass ratios as a function of the separation. }
  \label{fig:brill_lindquist_area}
\end{figure*}

Fig.~\ref{fig:brill_lindquist_area} shows the areas of the four
horizons as functions of $d$.  The apparent horizon is the easiest to
understand in terms of the irreducible mass
$M_{irr} = \sqrt{{\rm Area}/16\pi}$.  Since the black holes have zero
angular momentum, $M_{irr}$ is also the correct mass that one should
assign to the black hole \cite{Dreyer:2002mx}.  As discussed earlier
(see also \cite{Schnetter:2006yt}), the difference $M_{ADM} - M_{irr}$
is a measure of excess radiative energy present between the horizon
and spatial infinity.  As we make $d$ smaller and eventually set it to
zero, we have just a single black hole and in fact we recover a slice
of the Schwarzschild spacetime (in isotropic coordinates).  For a
Schwarzschild black hole, since it is globally static, there is
clearly no radiation and the horizon mass must equal the ADM mass.
The horizon area is then just $16\pi M_{irr}^2 = 16\pi M_{ADM}^2$.
Since we set $M_{ADM}=1$, we expect that as $d\rightarrow 0$,
${\rm Area} \rightarrow 16\pi \approx 50.265$ which is what is found
in Fig.~\ref{fig:sub_area1} for all the mass ratios.

For the inner and the larger individual MOTS shown in
Figs.~\ref{fig:sub_area2} and \ref{fig:sub_area3}, the areas are seen
to increase monotonically as $d$ is reduced.  Clearly, just when the
common horizon is formed, the inner MOTS and the AH coincide and their
areas must necessarily agree.  The inner MOTS area increases more
rapidly and eventually has a larger area than the apparent horizon.
The smaller horizon initially has a much smaller area than any of the
other marginal surfaces.  For $d\rightarrow \infty$, its irreducible
mass must agree with the bare mass $m_1$. Thus, for a mass ratio
$q = m_2/m_1$, we must have $M_{irr} \approx 1/(1+q)$ which implies
that its area is approximately $16\pi/(1+q)^2$. Thus, the area is
smaller for the more asymmetric system.  For the larger MOTS, the same
argument shows that the area must be $16\pi q^2/(1+q)^2$ for large
$d$.  This is larger for more asymmetric systems.

For very small $d$, the area of the smaller MOTS increases very
rapidly and in fact the ``small'' black hole ends up having the
largest area.  This can be understood by recalling that the punctures
are in fact asymptotically flat regions by themselves.  As the
individual horizon nears the puncture, it is in fact moving towards
asymptotic infinity at the other end of the Einstein-Rosen
bridge.  It is thus not surprising that its area increases rapidly for
very small $d$.

\begin{figure*}
  \begin{subfigure}{.45\textwidth}
    \centering
    \includegraphics[width=\textwidth]{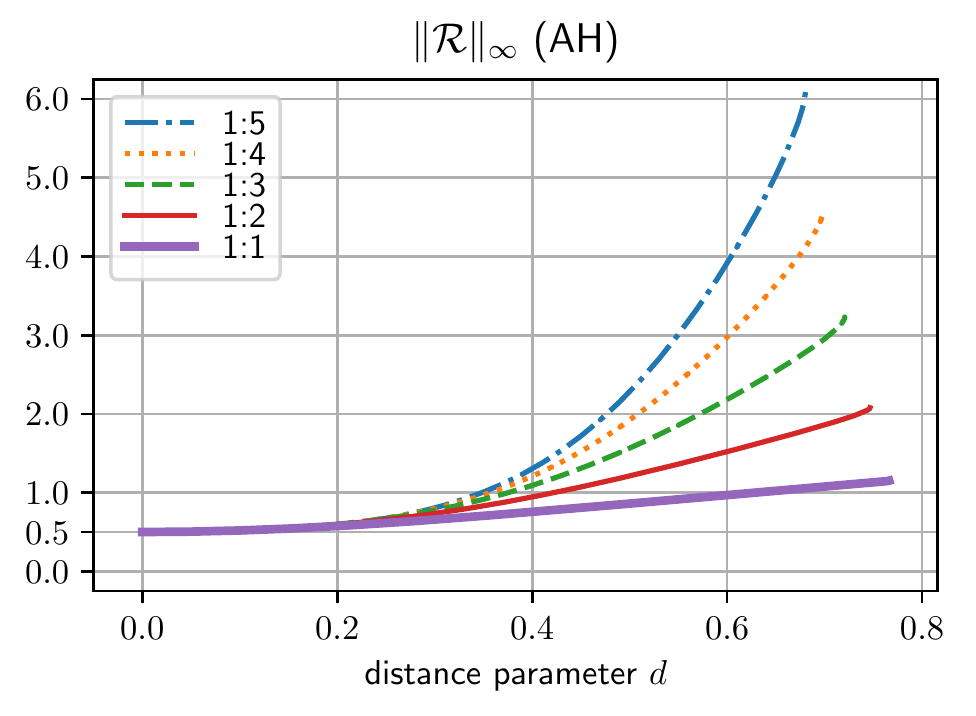}
    \caption{The maximum of the Ricci scalar on the AH.  For
      $d\rightarrow 0$ this approaches $0.5$, as expected for a
      ``round'' sphere. }
    \label{fig:sub_ricci1}
  \end{subfigure}
  \hfill
  \begin{subfigure}{.45\textwidth}
    \centering
    \includegraphics[width=\textwidth]{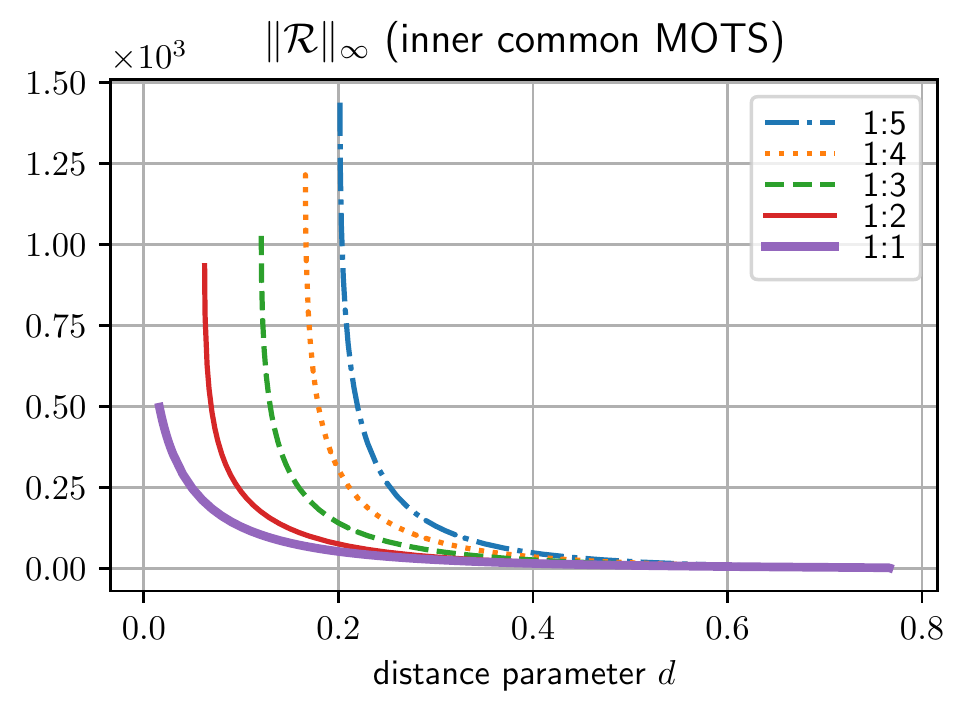}
    \caption{The inner common horizon becomes extremely distorted as
      $d\rightarrow 0$. This shows up as the sharp increase in the
      Ricci scalar right up to the point where the MOTS disappears. }
    \label{fig:sub_ricci2}
  \end{subfigure}
  \hfill  
  \begin{subfigure}{.45\textwidth}
    \centering
    \includegraphics[width=\textwidth]{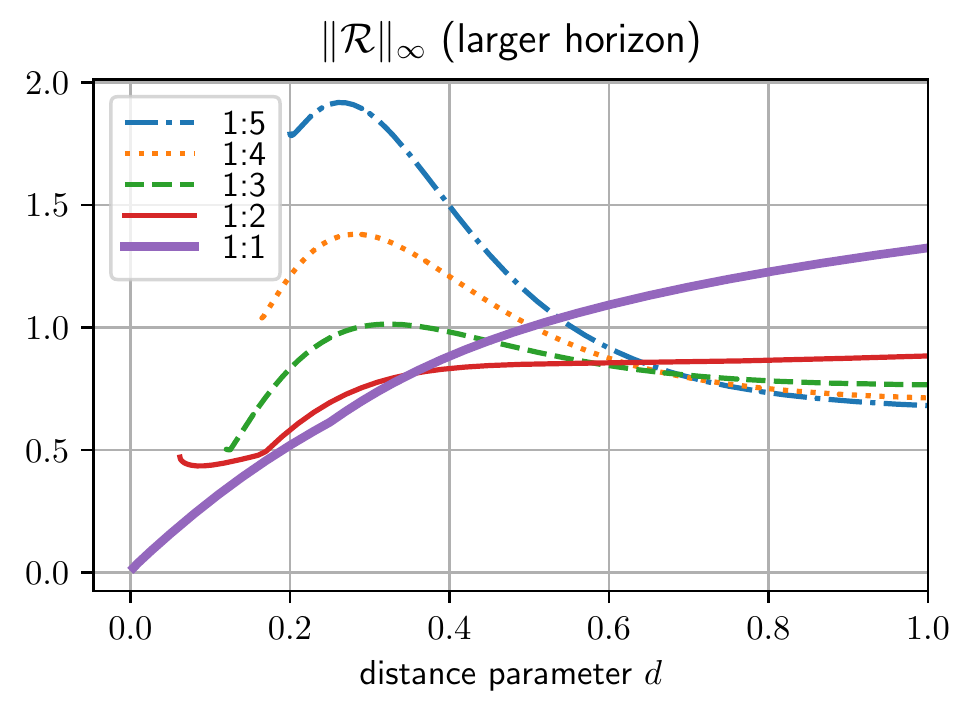}
    \caption{The Ricci scalar of the larger horizon.  The
      discontinuous behavior of the slope of some of the curves is
      because there are multiple peaks, and it is possible for the
      maximum to go from one peak to the other.  }
    \label{fig:sub_ricci3}
  \end{subfigure}
  \hfill
  \begin{subfigure}{.45\textwidth}
    \centering
    \includegraphics[width=\textwidth]{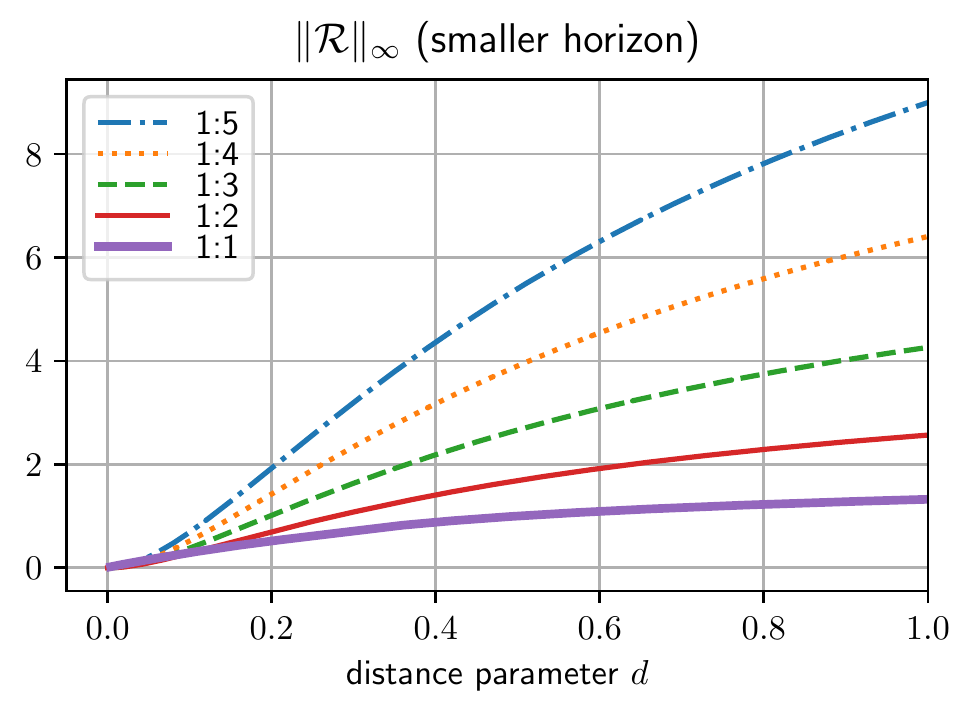}
    \caption{For $d\rightarrow 0$ the smaller MOTS moves towards
      asymptotic infinity which is why its Ricci scalar vanishes
      asymptotically. The equal mass line is of course the same as the
      equal mass line in the previous panel
      Fig.~\ref{fig:sub_ricci3}.}
    \label{fig:sub_ricci4}
  \end{subfigure}
  
  \caption{The maximum value of the Ricci scalar on the various MOTSs
    in BL data with different mass ratios as a function of the
    separation.  The purpose this plot is to give a rough quantitative
    idea about the distortions of the various marginal surfaces.}
  \label{fig:brill_lindquist_max_ricci}
\end{figure*}

The next set of plots shows the maximum value of the intrinsic Ricci
scalar on the horizons.  As before, the apparent horizon is easy to
understand.  For a ``round'' sphere of radius $R$ in Euclidean space,
the scalar curvature is $\mathcal{R} = 2/R^2$. Following the same
argument as above for the area we get
$\mathcal{R} = 8\pi/{\rm Area} = 8\pi/(16\pi M_{irr}^2) \rightarrow
0.5$ as $d\rightarrow 0$.  This is confirmed in
Fig.~\ref{fig:sub_ricci1}.  Similarly, the curvature of the smaller
MOTS, which we have seen approaches asymptotic infinity and infinite
radius, is $\mathcal{R} = 2/R^2 \rightarrow 0$ as also seen in
Fig.~\ref{fig:sub_ricci4}.  For the inner MOTS, we have seen already
in Fig.~\ref{fig:sub6} that it is highly distorted and is almost
pinching off at its neck.  Thus, we expect increasingly large
curvature at the neck which is what is seen in
Fig.~\ref{fig:sub_ricci2}.  The behavior of the larger individual MOTS
shown in Fig.~\ref{fig:sub_ricci3} shows an interesting maximum for
which we have no obvious explanation.

\begin{figure}[htpb]
    \includegraphics[width=0.45\textwidth]{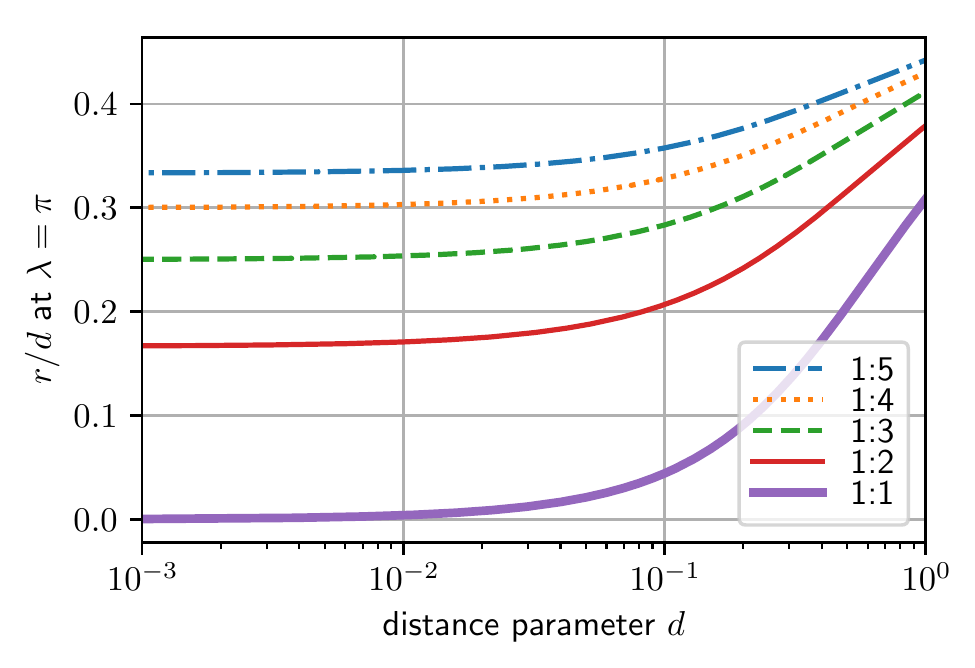}
    \caption{The ratio of the coordinate radius of the smaller MOTS and the
      puncture separation $d$, as $d$ approaches 0.  }
    \label{fig:r_over_d}
\end{figure}

We can now postulate a scenario for the final fate of the remaining
MOTS corresponding to the smaller black hole as $d\rightarrow 0$.  For
the exterior spacetime, in the ``close-limit'' approximation
\cite{Price:1994pm}, for small $d$ we should be able to express the
exterior spacetime as a perturbation of the Schwarzschild solution.
Thus we expect that the interior spacetime should also approach a
slice of Schwarzschild in this limit.  Since the coordinate radius of
the MOTS vanishes asymptotically, it is plausible that the MOTS
vanishes ``into'' the puncture, after which the puncture itself ceases
to exist.  Then we are just left with the other puncture so that
asymptotically we have just a slice of Schwarzschild spacetime.  This
scenario could go wrong if in fact the coordinate radius $r$ of the
MOTS does not decrease sufficiently rapidly with $d$, and the MOTS
intersects the other puncture.  Note that the proper distance will not
be useful here since the proper distance from the puncture is always
infinite.  The question therefore is what happens to the ratio $r/d$
as $d\rightarrow 0$.  This is shown in Fig.~\ref{fig:r_over_d}.
Except for the case of equal masses, we see that $r/d$ asymptotes to a
constant value less than unity which supports the scenario outlined
above.

\section{Stability and existence}
\label{sec:mainresults}

The previous section has illustrated some general properties of the
different marginal surfaces for various values of the mass ratio.  We
have seen cases when the inner and larger marginal surfaces are not
found or are highly distorted.  The fundamental question of their
existence and stability has been posed but we do not yet know why
they cease to exist in certain cases.  Are marginal surfaces
inherently ill-behaved or can these properties be understood in terms
of predictable and regular quantities?  In this section we answer
these questions and show that there is a deep link between existence
and stability.  We shall start by trying to find the various MOTSs
numerically as exhaustively as possible.  Later we shall turn to the
stability parameter to explain the failure to find some of these
marginal surfaces.

We have already demonstrated the stability of the apparent horizon and
the instability of the inner common MOTS in Fig.~\ref{fig:CE}.  Recall
that the stability of a MOTS is governed by the spectrum of the
stability operator defined in Eq.~\ref{eq:stability_operator}.  In
particular, positivity of the principal eigenvalue guarantees smooth
time evolution.  The stability parameter is defined to be the
principal eigenvalue.

\begin{figure}[htpb]
    \includegraphics[width=0.45\textwidth]{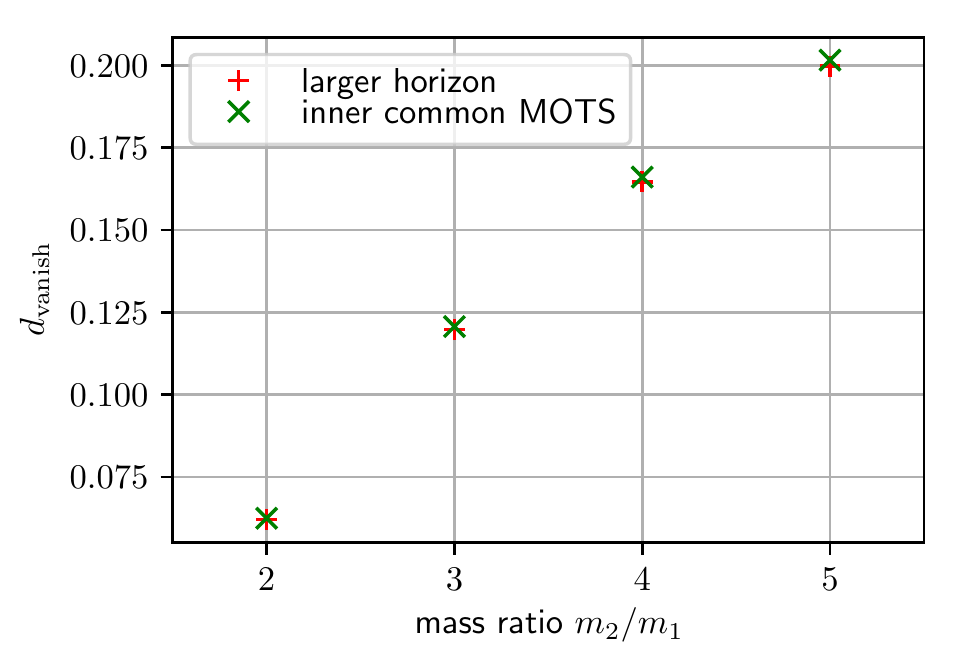}
    \caption{The values of $d$ for which the inner and larger MOTSs
      vanish for different mass ratios.  The difference between the
      values for each mass ratio are small (but still much larger than
      numerical error).  This is discussed in more detail in the next
      section (see Fig.~\ref{fig:stability_bot_inner_r4} below).  The equal
      mass case is qualitatively different and is not shown in this
      plot. }
    \label{fig:dvanish}
\end{figure}

Since stability is connected with existence, we begin by investigating
the critical values of $d$ below which the inner and larger MOTSs,
$\mathcal{S}_{in}$ and $\mathcal{S}_{large}$ respectively, cease to
exist.  Since the apparent horizon and smaller individual MOTS exist
for all $d$ as we have seen, the question of existence is more
relevant for the inner and larger MOTS.  Fig.~\ref{fig:dvanish} shows
the critical values of $d$, denoted $d_{vanish}$ for different mass
ratios. The inner MOTS is seen to vanish just a little bit before the
larger individual MOTS, i.e. the individual MOTS persists for slightly
smaller values of $d$, and this is not a numerical artifact.  It is
not the case that these two surfaces touch each other just before
vanishing.  Referring again to the properties of minimal surfaces, if
they did touch and their tangent vectors were aligned, then they would
have had to coincide. In this case, the smaller MOTS acts as a barrier
which prevents the $\mathcal{S}_{in}$ and $\mathcal{S}_{large}$ from
touching.

\begin{figure}[htpb]
    \includegraphics[width=0.45\textwidth]{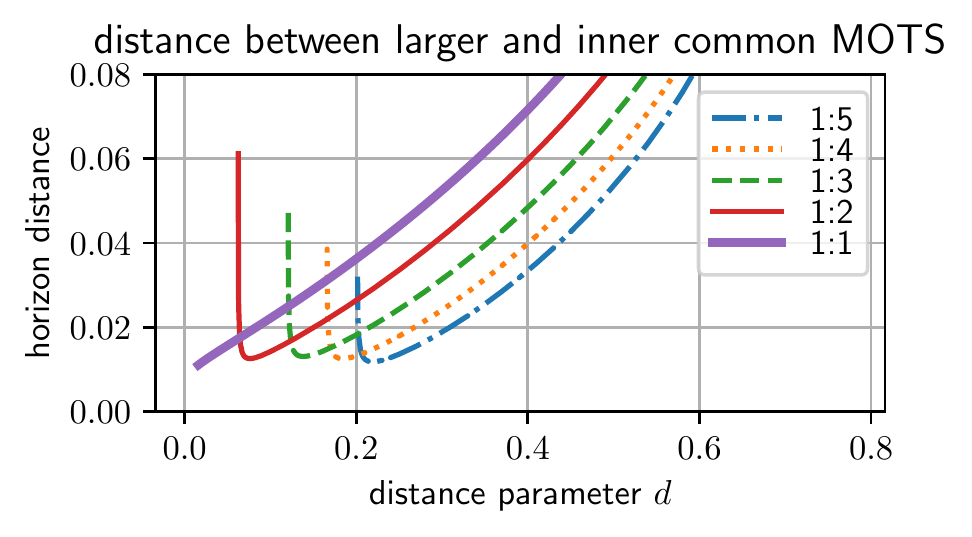}
    \caption{The proper distance between $\mathcal{S}_{inner}$ and
      $\mathcal{S}_{large}$ as a function of $d$.  As $d$ is
      decreased, the proper distance decreases initially as one might
      expect. Surprisingly, this does not continue and the curve turns
      over, and the proper distance is increasing just before
      $\mathcal{S}_{inner}$ vanishes.  This shows that
      $\mathcal{S}_{inner}$ and $\mathcal{S}_{large}$ remain well
      separated from each other. The equal mass case is qualitatively
      different and in the limit of equal masses we must
      asymptotically get the purple curve.  The calculations become
      more expensive as $q\rightarrow 1$ and it is not entirely
      straightforward to understand this limit numerically.  }
    \label{fig:properdistance}
\end{figure}

To show this conclusively, we can compute the proper distance between
$\mathcal{S}_{in}$ and $\mathcal{S}_{large}$ at their closest points
of approach, namely along the negative z-axis; see
Fig.~\ref{fig:properdistance}.  This shows that the two surfaces never
touch and are in fact well separated just before they disappear
compared to our numerical resolution.  

A search for constant expansion surfaces in the relevant region
reveals an upper limit on the value of the expansion.  As mentioned
earlier in Sec.~\ref{subsec:massratios}, we find that we do not have
any surfaces with $\Theta_{(\ell)} > c$ for some negative constant
$c$.  To show this, we start with the AH which has of course zero
expansion.  Then we proceed in the inward direction finding surfaces
with increasingly negative expansions.  We can parameterize these
surfaces also using their areas and we can plot the expansion as a
function of the area.  We expect that as we go inwards, we should
initially have the expansion decreasing and becoming more negative.
If the inner common MOTS exists, then this expansion must eventually
start to increase.  If it increases to zero, then we have found the
inner common MOTS and we should generally be able to extend the
expansion to positive values.  Taking the area as a function of the
expansion, we should expect to find the MOTSs appear as local minima.
As explained earlier, the constant expansion surfaces do not form a
regular foliation and can intersect each other.

\begin{figure}[htpb]
    \includegraphics[width=0.45\textwidth]{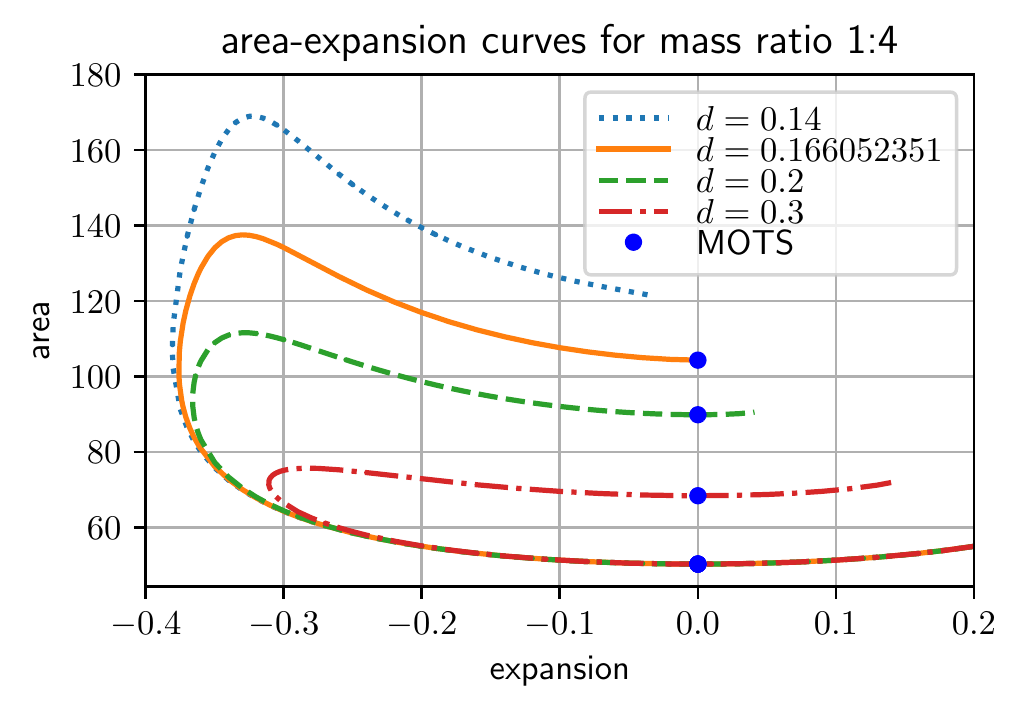}
    \caption{Constant expansion surfaces parameterized by the value of
      the expansion and the area for mass ratio 1:4, for different
      values of $d$.  See text for explanation. }
    \label{fig:area_expansion}
\end{figure}

The results of this search for constant expansion surfaces are shown
in Fig.~\ref{fig:area_expansion}.  In this figure, the blue dots
represent the MOTSs.  The lowest blue dot is the AH. The curves with
the different distance parameters do not actually coincide at that
point, but for the values of $d$ that we have chosen, the areas are
very close to each other and cannot be distinguished by eye on the
plot (see also the curve for mass ratio 1:4 in
Fig.~\ref{fig:sub_area1}). Going inwards from the AH corresponds to
moving leftwards from the AH and the curves for different $d$ can soon
be differentiated. All the curves eventually turn over and the
expansion starts to increase towards 0.  However, not all the curves
actually reach 0, e.g. for $d=0.14$ the curve stops well before that.
Thus, for $d=0.14$, not only are we unable to find surfaces of zero
expansion, we also cannot find any surfaces with $\Theta_{(\ell)} > c$
for some \emph{negative} number $c$.  At the critical value, the curve
stops precisely at 0 and for larger $d$ the curves extend to positive
expansions.  In each case, the MOTSs all occur at local minima of the
area.

Based on this detailed numerical study, we are led to suspect that the
inner and larger MOTS simply cease to exist below critical values of
$d$.  Since it is inherently difficult to show that something does not
exist, is this a limitation of our numerical method? Would we continue
to find these marginal surfaces if only we looked even more carefully?

\subsection{The stability parameter}

To answer this question, we turn now to the stability operator defined
in Sec.~\ref{subsec:definitions}.  As mentioned there, the sign of the
principal eigenvalue is important.  The positivity of this principal
eigenvalue implies stable time evolution and should thus also be
relevant to the question of whether the surfaces can be found in
Cauchy data.  We denote the principal eigenvalue as the
\emph{stability parameter}.  Before presenting numerical results, it
is useful to briefly describe what we might expect theoretically, and
in particular what kind of MOTSs are expected to have negative
stability parameter.  It can be shown that for vacuum time-symmetric
data, the stability operator of Eqs.~(\ref{eq:stability_operator}) and
(\ref{eq:stability_operator_bl}) can be written as
\begin{equation}
  Lf = -\Delta f + \frac{\mathcal{R}}{2} f \,.
\end{equation}
As before, $\mathcal{R}$ is the intrinsic Ricci scalar on
$\mathcal{S}$.  The spectrum of the Laplacian on a general distorted
sphere can be very complicated and from a full knowledge of the
spectrum we can infer some geometrical properties of $\mathcal{S}$
such as its area and its genus (see
e.g. \cite{Minakshisundaram:1949xg,Kac:1966xd}).  Asymptotic
properties of the spectrum are also known but here we shall need the
lower eigenvalues.  Regarding the MOTS stability operator itself,
Jaramillo has studied its spectrum for stationary axisymmetric
horizons leading to an interesting physical interpretation as a
``pressure operator'' \cite{Jaramillo:2013rda,Jaramillo:2014oha}.  The
stability operator is also related to extremality
\cite{Jaramillo:2012zi}.  Numerical studies should be able to extend
these results and lead to a better understanding of the stability
operator in generic situations.

Some obvious simplifications are possible in special cases.  For a
``round'' 2-sphere, with $\mathcal{R} = 2/R^2$, the principal
eigenvalue is simply
\begin{equation}
  \label{eq:stability_round}
  \Lambda_0 = \frac{1}{R^2} = \frac{1}{4M_{irr}^2}\,.
\end{equation}
This will be relevant for the AH as $d\rightarrow 0$, and for the
individual horizons as $d\rightarrow \infty$.  Thus, ``round'' spheres
are always stable and it follows then that the inner MOTS cannot be
spherically symmetric.  The higher eigenvalues are easily obtained from the
spectrum of the Laplacian on a sphere of radius $R$, i.e.
$\Lambda_n = [1+n(n+1)]/R^2, n=0,1,2,3,\ldots$, with multiplicity $2n+1$.

More generally in the absence of any symmetries, from the
Rayleigh-Ritz formula, if $\psi$ is a square integrable function on
$\mathcal{S}$:
\begin{equation}
  \label{eq:approx_stability}
  \Lambda_0 = \inf_{\psi}\int_\mathcal{S} \left(||\partial\psi||^2 +  \frac{1}{2}\mathcal{R}\psi^2\right) \,d^2V\,.
\end{equation}
Thus, we are more likely to get instabilities when there are
significant regions of negative curvature on $\mathcal{S}$; in fact,
there must be regions of negative curvature for unstable marginal
surfaces.  The work of
\cite{Andersson:2005gq,Andersson:2007fh,Andersson:2008up} also uses
the Rayleigh-Ritz formula, albeit a generalized version valid for the
non-self adjoint case.

\begin{figure*}
  \begin{subfigure}[ht]{.45\textwidth}
    \centering
    \includegraphics[width=\textwidth]{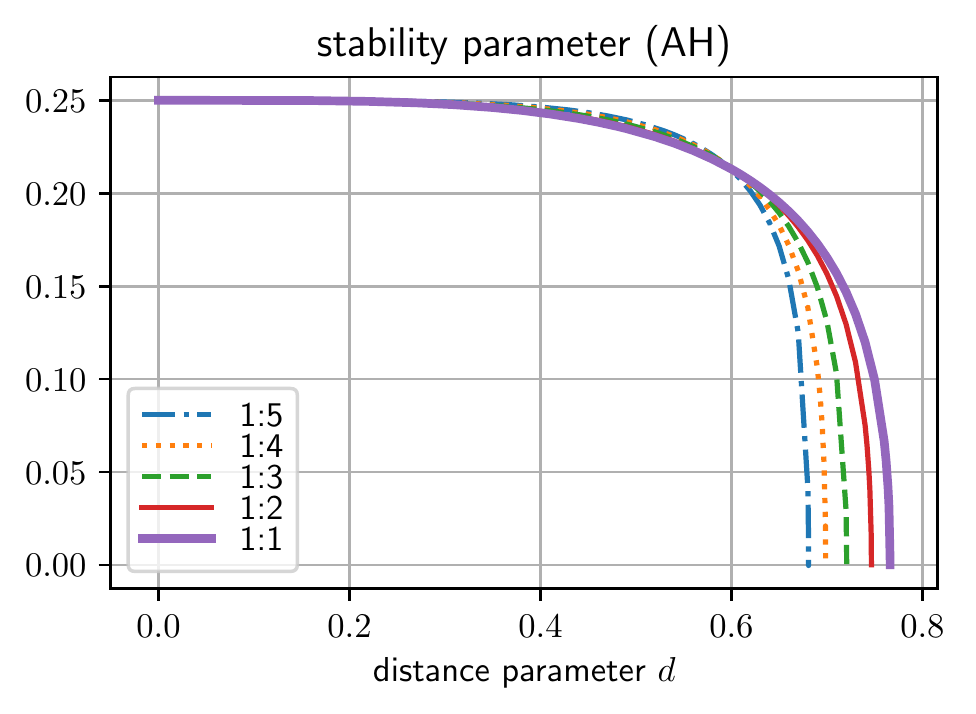}
    \caption{The stability parameter for the AH as a function of
      $d$. For all mass ratios, this approaches 0.25 for small $d$
      consistent with Eq.~(\ref{eq:stability_round}) and
      $M_{irr}\rightarrow 1$. }
    \label{fig:sub_stability1}
  \end{subfigure}
  \hfill
  \begin{subfigure}[ht]{.45\textwidth}
    \centering
    \includegraphics[width=\textwidth]{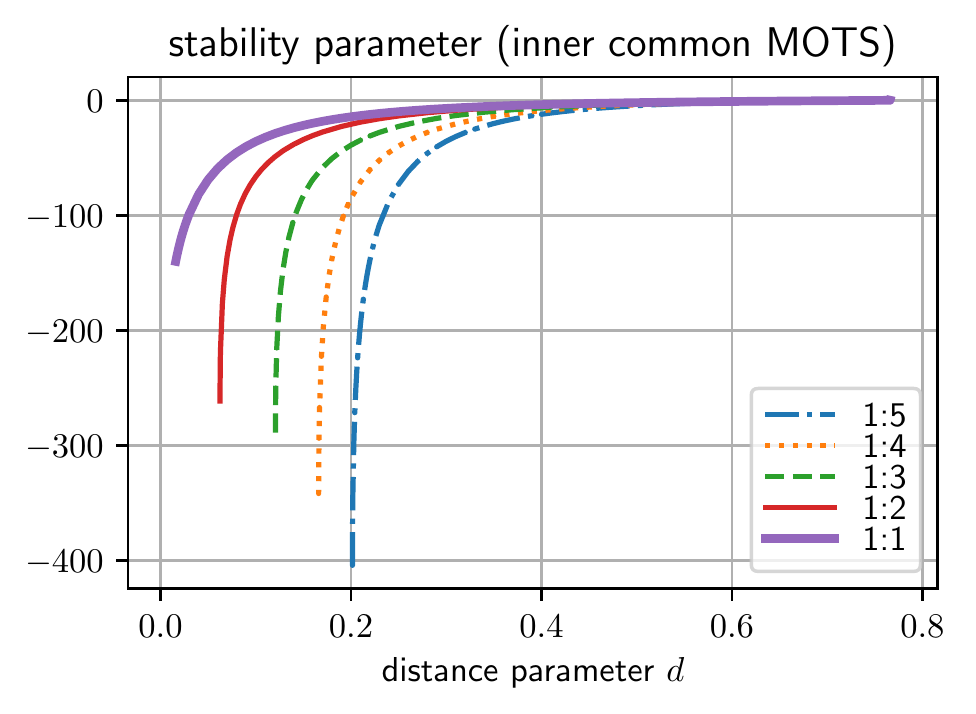}
    \caption{Stability parameter for the inner common MOTS.  This is
      initially zero when the MOTS is formed and decreases
      monotonically as $d$ is decreased further. We need to go beyond
      the principal eigenvalue to understand the existence of this
      MOTS. }
    \label{fig:sub_stability2}
  \end{subfigure}
  \hfill
  \begin{subfigure}[ht]{.45\textwidth}
    \centering
    \includegraphics[width=\textwidth]{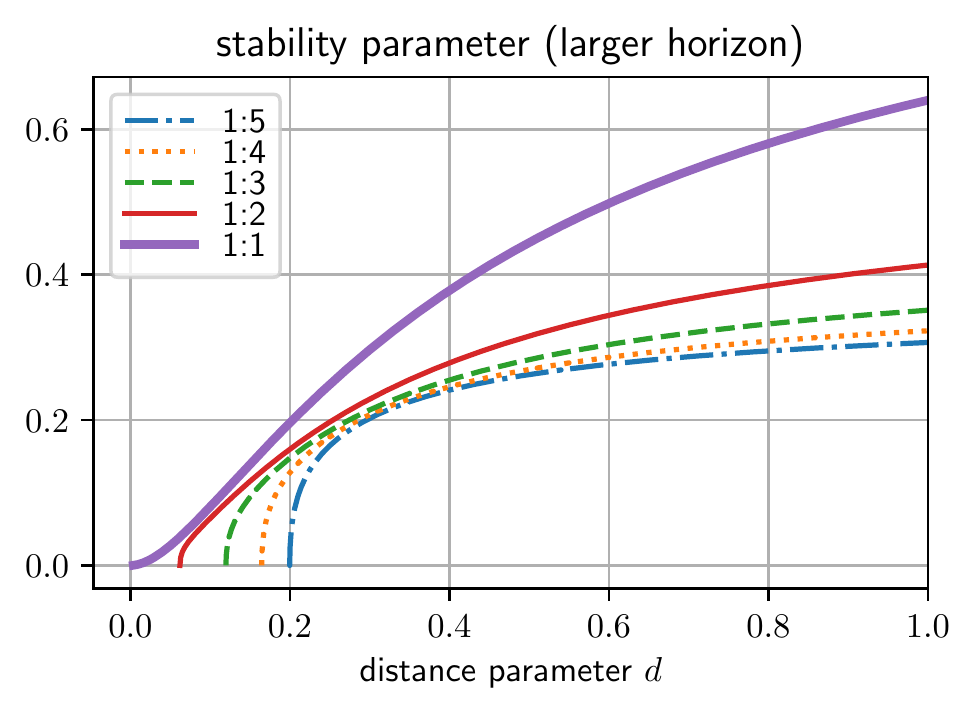}
    \caption{The stability parameter of the larger MOTS. This vanishes
      precisely when it can no longer be found numerically. }
    \label{fig:sub_stability3}
  \end{subfigure}
  \hfill
  \begin{subfigure}[ht]{.45\textwidth}
    \centering
    \includegraphics[width=\textwidth]{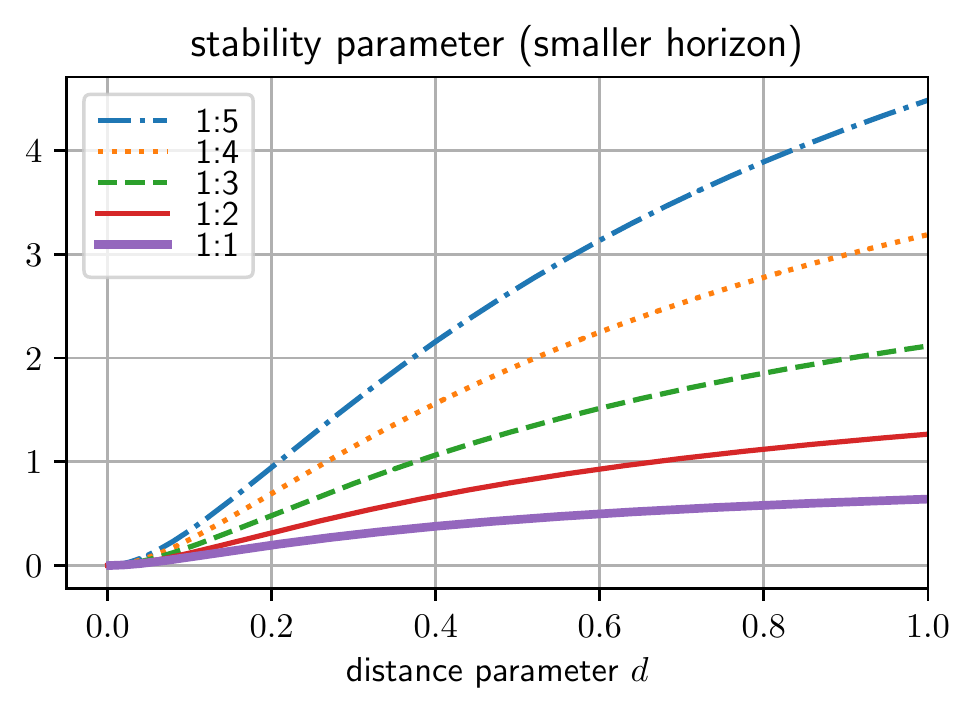}
    \caption{The stability of the smaller individual MOTS.  This
      vanishes asymptotically for small $d$ consistent with the
      scenario sketched out at the end of
      Sec.~\ref{subsubsec:massratios}.  For large $d$, following
      Eq.~\ref{eq:stability_round}, the stability parameter must
      approach $(1+q)^2/4$.}
    \label{fig:sub_stability4}
  \end{subfigure}
  
  \caption{The stability parameter for MOTSs in
    BL data with different mass ratios as a function of the
    separation. }
  \label{fig:brill_lindquist_stability}
\end{figure*}

We now present our numerical results for the stability parameter.
Plots of the stability parameter are shown in
Figs.~\ref{fig:brill_lindquist_stability} for all the four kinds of
marginal surfaces as a function of $d$.  These plots have several very
interesting features which we now discuss.

First, the apparent horizon is always seen to be stable as expected in
Fig.~\ref{fig:sub_stability1}. When it is first formed, its stability
parameter is zero but it rapidly increases before leveling off.  Its
asymptotic value is 0.25, consistent with the above argument for
$\Lambda_0$ since its irreducible mass approaches unity.  Similarly,
in Fig.~\ref{fig:sub_stability4} the smaller MOTS remains stable as it
grows, and its stability parameter vanishes asymptotically (consistent
again with it vanishing in the limit $d\rightarrow 0$).

\begin{figure}[htpb]
    \includegraphics[width=0.45\textwidth]{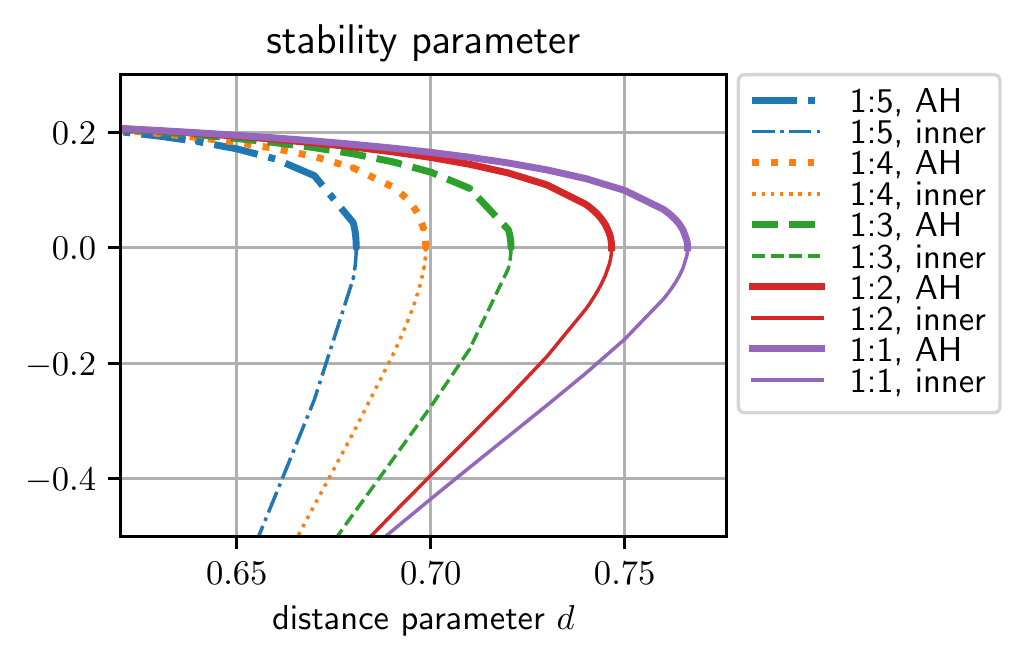}
    \caption{The stability parameters of the apparent horizon and the
      inner common MOTS near the point of formation for different
      mass-ratios. We see clearly the formation and bifurcation of the
      marginal surfaces into a stable and unstable branch. }
    \label{fig:stability_AH_inner}
\end{figure}

The other two horizons are particularly
interesting.
The stability parameter for the larger individual MOTS is positive,
and it ceases to exist exactly when the stability parameter
vanishes. This shows that the absence of this MOTS below a critical
value of $d$ is not an accident and is not a limitation of our
numerical method.

Now we turn to the inner common MOTS.
It is born with zero stability parameter
and, of course, it coincides with the apparent horizon at birth.
The stability parameter is strictly negative for smaller values of $d$,
and it decreases monotonically as $d$ is decreased; see
Fig.~\ref{fig:sub_stability2}. A close-up of the regime when the
common MOTSs are just formed is shown in
Fig.~\ref{fig:stability_AH_inner}, demonstrating the bifurcation of
the common MOTS into two branches.

\begin{figure}[htpb]
    \includegraphics[width=0.45\textwidth]{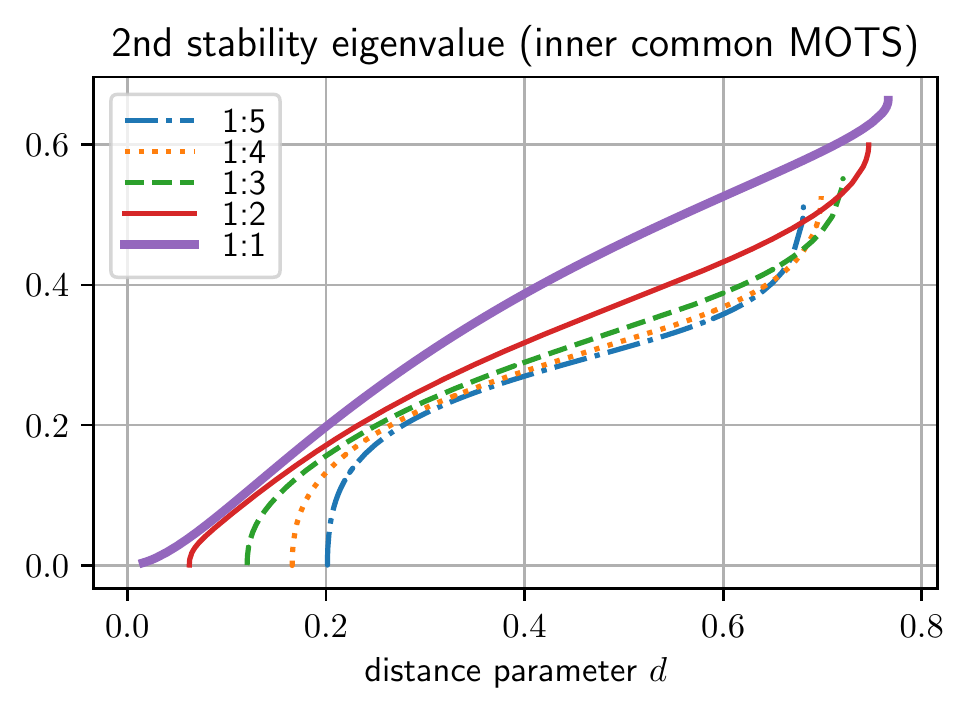}
    \caption{The second eigenvalue of the stability operator for the
      inner horizon for different mass ratios.  This MOTS ceases to
      exist precisely when the second eigenvalue vanishes.  As in many
      of the other plots, the equal mass ratio case is qualitatively
      different in that the individual MOTSs continue to exist and
      become unstable only asymptotically as $d\rightarrow 0$. This
      limit is however not easy to explore numerically.}
    \label{fig:stability2_inner}
\end{figure}

Does the existence of this MOTS,
for which the stability parameter is always negative and decreasing,
cast doubt on the relevance of the stability operator?  To investigate
this, we go back to the work of Andersson et al.
\cite{Andersson:2005gq,Andersson:2007fh,Andersson:2008up}. The
important issue is whether the stability operator is invertible,
i.e. we need to ensure that zero eigenvalue states are not
allowed. This is guaranteed automatically when the principal
eigenvalue is positive.  In the case when the principal eigenvalue is
negative, we must require that none of the higher eigenvalues
vanish. More specifically, since the principle eigenvalue is initially
zero and decreases monotonically, we need only investigate the second
eigenvalue.  When the MOTS is formed, it must be positive. The
question then is: does it decrease and does it ever vanish and then
become negative?  Fig.~\ref{fig:stability2_inner} shows the second
eigenvalue as a function of $d$ for the different mass ratios.  We see
that the eigenvalue vanishes exactly at the point when the inner
common MOTS vanishes.  Again, this supports our claim that the MOTS
does not exist for small $d$, and that its disappearance is not merely
a limitation of our numerical method. More importantly, we see again
the importance of the stability operator to the question of stability
and existence of marginal surfaces.

\begin{figure}[htpb]
    \includegraphics[width=0.45\textwidth]{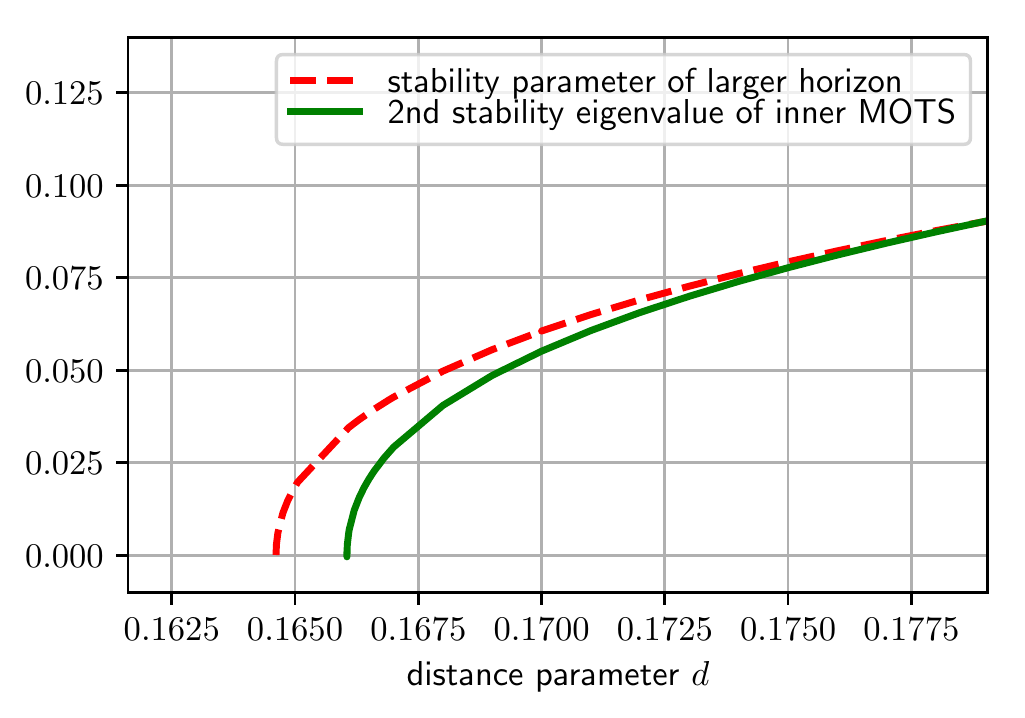}
    \caption{The second eigenvalue of the stability operator for the
      inner common MOTS and the principal eigenvalue for the larger individual
      horizon for mass ratio 1:4.  Both these curves have appeared in
      earlier plots, but here we show them close to where the surfaces
      cease to exist. We see that the eigenvalues both vanish and do
      so at different values of $d$.  The large MOTS persists somewhat
      longer than the inner common one. Similar results hold for the
      other mass ratios.}
    \label{fig:stability_bot_inner_r4}
\end{figure}

Finally, going back to the question of whether the inner and larger
MOTS vanish at different times (Fig.~\ref{fig:dvanish}), we can look
at when the stability parameter for the larger MOTS and the second
eigenvalue for the inner common MOTS vanish.
Fig.~\ref{fig:stability_bot_inner_r4} shows the principal eigenvalue
for the larger MOTS and the second eigenvalue for the inner common
MOTS.  Both of these eigenvalues vanish but at different values of
$d$. This confirms again that the inner common MOTS vanishes before
the larger individual MOTS.

\section{The mass multipoles}
\label{sec:multipoles}

The geometric mass multipole moments $I_n$ have been defined
previously in Eq.~\ref{eq:massmultipoles}.  We can apply this to any
of the MOTS that we have found, but it is the most interesting to
calculate them for the apparent horizon.  As described in
\cite{Gupta:2018znn}, these moments approach their final stationary
values (in general corresponding to a Kerr black hole), and it is of
interest to calculate the rate at which they decay.  In the present
case, the ``final'' black hole is Schwarzschild and $d$ is a proxy for
time. Thus, except $I_0$ (which is a geometric invariant), we expect
all the $I_n$ to vanish asymptotically as $d\rightarrow 0$.

\begin{figure}[h]
    \includegraphics[width=0.45\textwidth]{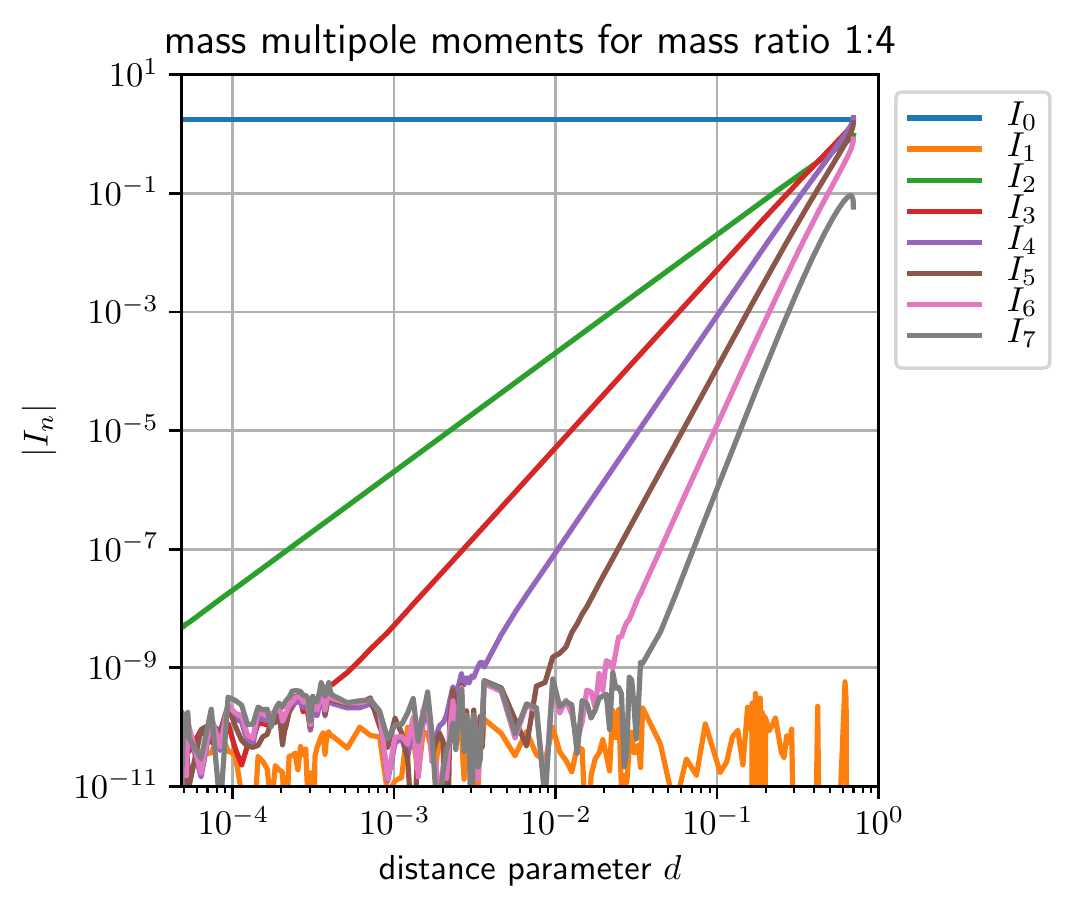}
    \caption{Mass multipoles $|I_n|$ of the apparent horizon as a
      function of the distance. Values below $10^{-9}$ are in the
      numerical noise.  This plot is for the 1:4 mass ratio
      configuration, however the falloff rates with $d$ do not depend
      on the mass ratio.  The first moment $I_0=\sqrt{\pi}$ is a
      geometric invariant, and the others falloff as $I_n\sim d^n$.  }
    \label{fig:multipoles}
\end{figure}
Fig.~\ref{fig:multipoles} shows the mass multipole moments for the 1:4
configuration as a function of $d$.  The lowest moment $I_0$ is a
geometric invariant, the integral of the Ricci scalar over
$\mathcal{S}$.  The Gauss-Bonnet theorem shows that $I_0 = \sqrt{\pi}$
for a sphere.  We see next that the mas dipole $I_1$ vanishes.  This
is true generally as shown in \cite{Ashtekar:2004gp}, with the
physical interpretation being that the invariant coordinates
automatically place us in the center-of-mass frame of the horizon.
For all $n\geq 2$, Fig.~\ref{fig:multipoles} shows that the $I_n$
fall off as power laws as $d\rightarrow 0$ and it turns out that
this fall-off rate is \emph{independent} of the mass ratio. This is
similar to what one expects in a time evolution. When a distorted
horizon is initially formed, its initial configuration and which
moments are excited depends on the data which produced the horizon.
Thus, in a binary black hole system, this would depend on the mas
ratio, spin configurations etc.  However, the approach to equilibrium
has universal properties and this is reflected in the fall-off of the
$I_n$.  A simple numerical fit of the numerical data for mass ratio
1:4 and other mass ratios gives:
\begin{equation}
  I_n \propto d^n\,,\quad n \geq 2\,.
\end{equation}
We have verified this for $n\leq 7$.  Whether this behavior carries
over to time evolutions remains to be seen.

\section{Conclusions}
\label{sec:conclusions}

The stability operator is known to be very important in mathematical
studies of marginally trapped surfaces.  Among other results, it shows
the link between stability and smoothness under time evolution.  It
also controls other properties of a MOTS in a given time slice, such
as its behavior as a barrier for completely trapped and untrapped
surfaces.  In this paper we have shown the importance of the stability
operator for understanding marginally trapped surfaces in numerical
calculations.  By monitoring the lowest eigenvalues of the stability
operator, we can effectively diagnose whether any problems might be
expected in the horizon finder or in the time evolution.  In the case
when the principal eigenvalue is already negative, then the second
eigenvalue must be considered.  The horizon will cease to exist when
this second eigenvalue vanishes.

In general, the stability operator is not self-adjoint.  Thus, apart
from the principal eigenvalue, all higher eigenvalues will be
complex. It is easier for them to avoid the origin even if their real
parts vanish.  For the inner horizon in generic cases when the
principal eigenvalue is already negative, it is not clear if there is
any reason for the eigenvalues to precisely vanish.  It might also
happen that instabilities can arise when the eigenvalues get
sufficiently close to 0.  Following up on the results presented here,
these questions will be investigated in forthcoming work.

Another important part of this paper is a new numerical algorithm and
its implementation for locating MOTSs capable of finding highly
distorted surfaces with no additional computational cost.  The method
is a modification of the commonly used \texttt{AHFinderDirect} and is
based on choosing a reference surface.  We have implemented a
pseudo-spectral scheme to represent the surface and we use a
Newton-Kantorovich method for solving the resulting non-linear PDE.
This implementation is at present valid in axisymmetry, but no
in-principle difficulty is foreseen for the full 3-dimensional case.
This will be incorporated into the Einstein Toolkit software and thus
available generally for black hole simulations.

We have applied this method to sequences of Brill-Lindquist data as
the separation between the puncture is decreased and a rich structure
of marginal surfaces is explored.  The distance parameter $d$ can be
pushed to 0 and the horizon finder and some of the inner marginal
surfaces become highly distorted.  Our horizon finder is able to
locate these with high accuracy.  The stability parameter works as
advertised: the larger individual horizon ceased to exist precisely
when its stability parameter vanishes.  The inner horizon is born with
zero stability parameter and it decreases monotonically as $d$ is
decreased.  The second eigenvalue thus becomes relevant and this MOTS
disappears exactly when this second eigenvalue vanishes.
Distorted as they are, the MOTSs are successfully tracked by our new numerical horizon finder
all the way to their disappearance due to losing physical stability.

Finally, we have found universal behavior (i.e. independent of the
mass ratio) of the mass multipole moments in the limit
$d\rightarrow 0$.  Forthcoming work will apply this horizon finder to
time evolutions.  One of the goals will be to find the fate of the
inner horizons in binary black hole spacetimes and to verify if this
universality in the approach to the final state still holds.

\acknowledgments

We thank Bruce Allen, Abhay Ashtekar, Jose-Luis Jaramillo, Istvan Racz
and Jeff Winicour for valuable discussions.  We are especially
grateful to Lars Andersson for pointing out the importance of the
invertibility of the stability operator to us.
O.B. acknowledges the National Science
Foundation (NSF) for financial support from Grant No. PHY-1607520.
The research was also supported by the Perimeter Institute for
Theoretical Physics. Research at Perimeter Institute is supported by
the Government of Canada through Industry Canada and by the Province
of Ontario through the Ministry of Research and Innovation.

\bibliography{mots}{}

\end{document}